\documentclass[twocolumn]{aastex631}

\newcommand{\Fdnu}{$F_{\Delta \nu} \ $}
\newcommand{\Fnmax}{$F_{\nu_{max}} \ $}
\newcommand{\aml}{$\alpha_{mlt} \ $}
\newcommand{\Dnu}{$\Delta \nu \ $}
\newcommand{\numax}{$\nu_{max} \ $}
\newcommand{\Teff}{$\text{T}_{\text{eff}}$ \ }

\usepackage{amsmath}
\usepackage{apjfonts} 
\usepackage{amsmath,amstext}
\usepackage[T1]{fontenc}
\usepackage{apjfonts} 
\usepackage[figure,figure*]{hypcap}
\usepackage{tablefootnote}

\begin{document}

\title{Testing the Breakdown of the Asteroseismic Scaling Relations in Luminous Red Giants}

\correspondingauthor{Amanda L. Ash}
\email{ash.172@buckeyemail.osu.edu}

\author[0000-0002-8674-9922]{Amanda L. Ash}
\affiliation{Department of Astronomy, The Ohio State University\\
140 West 18th Ave\\
Columbus, OH 43210, USA}

\author[0000-0002-7549-7766]{Marc H. Pinsonneault}
\affiliation{Department of Astronomy, The Ohio State University\\
140 West 18th Ave\\
Columbus, OH 43210, USA}

\author{Mathieu Vrard}
\affiliation{Observatoire de la Côte d’Azur, CNRS, Laboratoire Lagrange \\ 
Bd de l’Observatoire, CS 34229, \\
06304 Nice Cedex 4, France}
\affiliation{Department of Astronomy, The Ohio State University\\
140 West 18th Ave\\
Columbus, OH 43210, USA}

\author[0000-0002-7550-7151]{Joel C. Zinn}
\affiliation{Department of Physics and Astronomy, California State University, Long Beach\\
Long Beach, CA 90840, USA}



\begin{abstract}

Nearly all cool, evolved stars are solar-like oscillators, and fundamental stellar properties can be inferred from these oscillations with asteroseismology. Scaling relations are commonly used to relate global asteroseismic properties, the frequency of maximum power \numax and the large frequency separation \Dnu, to stellar properties. Mass, radius, and age can then be inferred with the addition of stellar spectroscopy. There is excellent agreement between seismic radii and fundamental data on the lower red giant branch and red clump. However, the scaling relations appear to breakdown in luminous red giant stars. We attempt to constrain the contributions of the asteroseismic parameters to the observed breakdown. We test the \numax and \Dnu scaling relations separately, by using stars of known mass and radius in star clusters and the Milky Way's high-$\alpha$ sequence. We find evidence that the \Dnu-scaling relation contributes to the observed breakdown in luminous giants more than the \numax relation. We test different methods of mapping the observed \Dnu to the mean density via a correction factor, \Fdnu\ and find a $\approx 1 - 3\%$ difference in the radii in the luminous giant regime depending on the technique used to measure \Fdnu. The differences between the radii inferred by these two techniques are too small on the luminous giant branch to account for the inflated seismic radii observed in evolved giant stars. Finally, we find that the \Fdnu correction is insensitive to the adopted mixing length, chosen by calibrating the models to observations of \Teff.

\end{abstract}


\section{Introduction}

Constraining the history and evolution of the Milky Way relies on accurate fundamental stellar properties. Fields such as galactic archaeology and stellar population synthesis rely on having precise measurements of mass, radius, and age for large populations of stars. Fortunately, we are in an exciting era where large data surveys are transforming our understanding of the galaxy. Spectroscopic surveys such as Galah and APOGEE have provided a wealth of chemical abundance data \citep{2022ApJS..259...35A, 2022MNRAS.517.5325H}. Photometric surveys like \textit{Gaia} have provided precise distances and magnitudes for over a billion Milky Way stars  \citep{2023A&A...674A...1G}. Red giant stars are of particular importance in these surveys as they are luminous and can be observed to large distances. In addition, they provide precise chemical abundance measurements. However, fundamental stellar properties are difficult to infer for large populations of red giant stars from classical HR diagram measurements. 

By adding time-domain astronomy, we are able to study and characterize the oscillations and variability of stars. Asteroseismology, or the study of these oscillations, has proven to be exceptionally precise at determining the masses and radii for stars in the core He-burning, or red clump (RC) phase, and on the lower red giant branch (RGB) (e.g. \citet{2011ApJ...743..143H}, \citet{2012ApJ...757...99S}, \citet{2016ApJ...832..121G}). By extracting fundamental properties from these oscillations, the frequency of maximum power, \numax, and the large frequency separation, \Dnu, we are able to infer masses and radii for large populations of stars using the asteroseismic scaling relations. However, we observe a breakdown in the scaling relations for more evolved red giant stars \citep{2019ApJ...885..166Z}. Constraining accurate stellar properties for luminous giant stars is of marked interest because these stars are detectable as resolved sources even in extragalactic systems, making them potent stellar population tracers. Despite these stars being bright, with luminosities $\gtrapprox 250 L_{\odot}$, and having high amplitude oscillations, there is an observed systematic inflation in the seismic radii of these stars of $\approx 9 \%$ relative to seismic-independent radii \citep{2019ApJ...885..166Z}. This breakdown may come from a number of sources. Luminous giant stellar oscillations are difficult to measure and interpret due to a small number of observationally accessible modes. One could also imagine that this breakdown is also inherent to the scaling relations. For example, in asymptotic red giant branch (AGB) stars, oscillation amplitudes become large enough that non-linear effects become important, so the standard scaling relations cannot adequately reproduce masses and radii, though one can develop new scaling relations \citep{2021MNRAS.500.1575T, 2023AJ....166..249H}. However, below the RGB tip, the oscillation amplitudes are smaller, meaning the stars are solar-like and the scaling relations should be applicable \citep{2013A&A...559A.137M}. In this paper, we test the scaling relations with independent masses and radii, and use stellar theory to study the impact of different methods for interpreting the oscillation pattern. 

\subsection{Joint Asteroseismic Scaling Relations}
Asteroseismology exploits the properties of sound waves propagating through a star to reveal information about the stellar interior. Waves generated by turbulence in the star are refracted by density gradients and can be trapped within the acoustic cavity of the star if conditions are right for the waves to be reflected at the surface. These trapped modes cause the star to oscillate at particular frequencies, causing the star's brightness to fluctuate over time. We can use Fourier analysis to construct a power spectrum from stellar light curves which can be described with spherical harmonics \citep{1991sia..book..401C}. 

There are two global properties that can be extracted from the resultant power spectrum, the large frequency separation, \Dnu, and the frequency of maximum power \numax. The large frequency separation is the distance between two adjacent radial mode in the power spectrum and the frequency of maximum power is the center of the smoothed excess power distribution. There is a third term, \Fdnu, that captures deviations from uniform spacings between adjacent modes. Using global properties rather than individual frequencies has several advantages. Global seismic parameters produce excellent agreement between fundamental stellar properties on the lower RGB and RC \citep{2024arXiv241000102P}.



\numax, \Dnu, and \Fdnu can be related to the stellar properties by making simplifying assumptions about the stellar structure \citep{1991ApJ...371..396B, 1995A&A...293...87K}. From asymptotic theory, \Dnu can be related to the mean stellar density as: 

\begin{equation}
    \Delta \nu^2 \propto \textlangle \rho \textrangle, 
\end{equation}

in the limit of large radial order, $n \gg 1$ \citep{1980ApJS...43..469T}. The Sun is typically used as a reference value. However, the structure of stars is not strictly homologous to the structure of the sun and the modes we are able to observe deviate from uniform spacing \citep{2011ApJ...742L...3W}. Therefore we apply a correction term to the observed \Dnu that is defined as: 

\begin{equation}\label{eq:Fdnu}
    \frac{(F_{\Delta \nu} \Delta \nu_{obs})}{\Delta \nu_{\odot}} = \bigg(\frac{(M/M_{\odot})}{(R/R_{\odot})^3}\bigg)^{1/2}, 
\end{equation}

\citep{2011ApJ...743..161W}. Note that the correction term as written in \citet{2011ApJ...743..161W} is written in terms of the mean stellar density, however it is analogous to the term as written in this work.

The \Fdnu term can be computed from the mean density given by stellar evolution models and depends on the stellar properties (e.g. composition and evolutionary state) as well as the chosen physics of the models. In addition, because this term captures deviations from uniform spacing, it provides more information about the frequency spectrum than an average of the frequency spacings across the spectrum does.

Alternatively, one could adopt a boutique modelling approach and apply corrections for surface effects to the individual frequencies in the power spectrum. This approach would account for shifts in mode frequencies due to improper modelling of the outer layers and using this framework can improve agreement between individual modelled and observed frequencies  \citep{dziembowski_paterno_ventura1988,monteiro_christensen-dalsgaard_thompson1996, kjeldsen_bedding_christensen-dalsgaard2008}. The effect of the surface correction on \Fdnu has been investigated recently in \citet{2023MNRAS.523..916L} who demonstrate how the surface effect corrections effect the agreement between seismic and parallactic radii for RGB stars up to $\approx 20 R_{\odot}$. Using surface corrections, they produce seismic radii that are smaller than parallactic radii by $2 - 6 \%$. This downward shift in seismic radii produces worse agreement between seismic and parallactic radii on the lower RGB than the \citet{2016ApJ...822...15S} correction terms, which represent a model-dependent technique, akin to that discussed previously. Additionally, this study found little evidence suggesting a radius-dependent effect, indicating that the surface effect corrections may be accounted for by a zero-point calibration term. Because the Li et al. (2023) method did not produce improved agreement with data, we instead assume that the surface effects are absorbed into the \Fnmax term presented in \citet{2024arXiv241000102P}.

The other component of the frequency spectrum, \numax, arises because waves will only be reflected if their wavelength is longer than the density scale height near the stellar surface. The pressure scale height is related to the surface gravity of the star. Therefore high frequency, short wavelength modes are not seen in stars with low surface gravities. For an isothermal atmosphere, \numax can be related to the surface gravity and temperature of a star as: 

\begin{equation}
    \nu_{max} \propto \frac{g}{T_{eff}^{1/2}}
\end{equation} 
\citet{1995A&A...293...87K}. 

However, this relationship is only formally valid for a strict set of physical assumptions. A relationship between the surface gravity and \numax also arises in a full physical model, but to form a generically applicable scaling with \numax one would need a predictive theory of turbulence and a detailed model of mode damping and excitations \citep{2011A&A...530A.142B}. While initial attempts appear promising, we lack such tools at present. Therefore, published work in practice treats the zero point of the \numax scaling relation as an adjustable parameter \Fnmax, to be calibrated empirically. From these equations, we can arrive at the joint asteroseismic scaling relations used to determine stellar mass and radius. The radius equation is given by: 

\begin{equation} \label{eq: radius scaling}
    \frac{R_*}{R_{\odot}} = \bigg(\frac{F_{\nu_{max}} \nu_{max}}{\nu_{max, \odot}}\bigg) \bigg(\frac{F_{\Delta \nu}\Delta \nu}{\Delta \nu_{\odot}}\bigg)^{-2} \bigg(\frac{\text{T}_{\text{eff}}}{\text{T}_{\text{eff}, \odot}}\bigg)^{1/2}, 
\end{equation}

and the mass relation by: 

\begin{equation} \label{eq: mass scaling}
    \frac{M_*}{M_{\odot}} = \bigg(\frac{F_{\nu_{max}} \nu_{max}}{\nu_{max, \odot}}\bigg)^{3} \bigg(\frac{F_{\Delta \nu}\Delta \nu}{\Delta \nu_{\odot}}\bigg)^{-4} \bigg(\frac{\text{T}_{\text{eff}}}{\text{T}_{\text{eff}, \odot}}\bigg)^{3/2}
\end{equation} 
\citet{2018ApJS..239...32P}. In \citet{2018ApJS..239...32P}, the \Fnmax term was treated as a scale factor inferred by adjusting the zero-point for the solar \numax.

Previous studies have used seismic-independent stellar parameters to test the accuracy of the asteroseismic scaling relations. \citet{2012ApJ...757...99S}, used \textit{Hipparcos} parallaxes to find good agreement between seismic distances and \textit{Hipparcos} parallaxes. \citet{2017ApJ...844..102H} compared asteroseismic radii to radii computed using parallax and luminosity information from \textit{Gaia DR1} for 2200 stars in the \textit{Kepler} field, and they found good agreement between \textit{Gaia DR1} radii and seismic radii up to a radius of $10 R_{\odot}$. Eclipsing binaries and open cluster stars have also been used as diagnostics to compare against asteroseismic properties (e.g. \citet{2016ApJ...832..121G} and \citet{2012MNRAS.419.2077M}). 

\citet{2019ApJ...885..166Z} used a larger asteroseismic sample and Gaia DR2 data, and found that the seismic radii agree with fundamental data to within $\approx 2\%$ on the lower giant branch. However in their sample, stars with seismic radii $\geq 30 R_{\odot}$, are systematically inflated by $\approx 9\%$ relative to the parallactic radii, which implies larger mass and age offsets. 

\subsection{Single Asteroseismic Scaling Relations}\label{sec:single scalings}

With independent masses and radii, we can test the ability of \numax and \Dnu to reproduce observed stellar properties separately. There are regimes in which measuring \numax or \Dnu are more difficult. In the luminous giant regime, there are few observable modes, making it more difficult to precisely centroid a power excess from the oscillation spectrum. Therefore, determining \numax may be difficult and it is advantageous to use the \Dnu scaling relation. In instances where there are low signal to noise ratios, it is difficult to characterize the frequency pattern, and therefore determining \numax is easier than determining \Dnu. In fact, \numax, but not \Dnu, has been reported for large numbers of red giants observed by the TESS satellite, and \numax alone could be a powerful diagnostic for the Roman Telescope Galactic Bulge Time Domain Survey \citep{2021ApJ...919..131H, 2015JKAS...48...93G}.

This latter case may have important implications on upcoming surveys including seismology measurements in the Galactic center using the Nancy Grace Roman Space Telescope \citep{2015JKAS...48...93G}. 

Other works have tested the efficacy of the single-asteroseismic scaling relations across the red giant branch. \citet{2011ApJ...743..143H} used a combination of stellar evolution models and \textit{Kepler} data for $\approx 1700$ RGB stars to test the \numax and \Dnu scaling relations and they found that the \numax and \Dnu scaling relations are in good qualitative agreement for lower RGB stars. Similar results were found in \citet{2012ApJ...760...32H} using interferometry up to a maximum radius of $12.5 R_{\odot}$. 

In this paper, we address limitations in the scaling relations, with particular attention to the luminous giant regime. We first focus on empirically testing the \numax and \Dnu scaling relations using open cluster data from NGC 6791 and NGC 6819 and the Milky Way's high-$\alpha$ sequence to get seismic independent mass checks. Here we use the Milky Way's high-$\alpha$ sequence as a "pseudo-cluster" of stars that are approximately coeval with a similar mean mass and age. The high-$\alpha$ sequence in the Milky Way has a strongly peaked mass distribution associated with the age of the thick disk, so we can use it as a benchmark group of stars with known radii and the same mean mass \citep{2021A&A...645A..85M}.  For NGC 6791 and NGC 6819, the RGB lifetime is short compared to the main sequence lifetime, so we fit for an isochronal mass to get a seismic-independent mass in the clusters. Our approach expands upon previous literature focusing on the single-parameter scaling relations as it uses new data from \textit{Gaia DR3}, and it investigates more evolved red giant stars than previously done. 

From a theoretical approach, we choose to focus on how matching stellar evolution models to observations affects \Fdnu. We test how different methods of measuring \Dnu in the theoretical frequency spectrum and assigning the appropriate model to a given star affect the agreement between the seismic and parallactic radius for the APOKASC3 sample. This methodology is similar to that of \citet{2023MNRAS.525.5540Z} who demonstrated that accounting for non-adiabatic effects improved the agreement between seismic and parallactic radii.

In Section \ref{sec:data}, we present the data from the APOKASC3 catalog and the \textit{Gaia} parallactic radii. We also present the parameters for selecting open cluster members and the high-$\alpha$ sequence. In Section \ref{sec: cluster methods} we discuss the methods used in the cluster component analysis and their results. In Section \ref{sec:model methods} we discuss the construction of the models and the methods for interpreting model results. Finally, in Section \ref{sec: conclusions} we summarize the conclusions and their implications.

\section{Data} \label{sec:data}

We begin by summarizing the surveys used for this paper. 

\subsection{Spectroscopic Measurements}

The temperatures and abundances we use in this paper come from APOGEE data release 17 (DR17) \citep{2022ApJS..259...35A, 2017AJ....154...94M, 2015AJ....150..173N, 2019PASP..131e5001W, 2017AJ....154..198Z}. APOGEE is an infrared spectroscopic survey that takes high-resolution spectra ($R \approx 22500$) for stars in the Milky Way galaxy. DR17 contains spectroscopic measurements for $657,000$  stars in the Milky Way galaxy. It provides elemental abundances for $18$ elements measured using the ASPCAP pipeline \citep{2016AJ....151..144G}. The temperatures measured from the APOGEE spectra are on an absolute temperature scale and are inferred using the infrared flux method \citep{2009A&A...497..497G}. 

\subsection{Asteroseismic Measurements}

The asteroseismic parameters in our data sample are measured from \textit{Kepler} light curves \citep{2010Sci...327..977B}. Originally designed for the detection of exoplanet transits, the \textit{Kepler} mission also obtained time series data on a large number of evolved red giant stars for which asteroseismic analysis has been performed \citep{2014ApJS..215...19P, 2018ApJS..239...32P, 2018ApJS..236...42Y}.

For this study, we use data from the APOKASC3 catalog which represents all stars with APOGEE spectra and \textit{Kepler} light curves \citep{2024arXiv241000102P}.
We use the APOKASC3 dataset because it has uniform, high quality data from \textit{Kepler}, APOGEE and \textit{Gaia}. The APOKASC3 catalog uses several methods to measure asteroseismic variables, with a calibration method that ties the asteroseismic scale to the fundamental scale. The total APOKASC3 sample contains $15809$ targets, $8969$ of which are shell H-burning, or RGB stars as deduced from asteroseismic analysis 
\citep{2024arXiv241103101V}. $1923$ of these stars are categorized as luminous giants with seismic radii $\geq 30 \ R_{\odot}$. 

APOKASC3 does not provide the average mass, radius, or \Fdnu measurements for stars with \numax values less than $1 \ \mu Hz$ because of the poor agreement between the joint scaling relations and fundamental data in this domain, which we are investigating. For these stars, we use the average asteroseismic measurements and our models to compute the mass, radius, and \Fdnu correction factor. 

\subsection{Parallactic Radii}\label{sec:parallactic radii}

The \textit{Gaia} mission provides precise parallax and proper motion measurements for 1.46 billion stars in the Milky Way galaxy \citep{2023A&A...674A...1G}. By pairing these precise parallax measurements, absolute temperatures, and photometry, one can infer the radius of the star via the Stefan-Boltzmann law. \citet{2019ApJ...878..136Z} uses this method to show a comparison between the asteroseismic radii in the APOKASC2 catalog and a seismic independent radius calculated using \textit{Gaia DR2} parallax information. The parallactic radii in this study are inferred using APOGEE spectroscopic temperatures, 2MASS $K_s$ band photometry, MIST isochrones, and \textit{Gaia}  DR3 parallaxes. For a more complete discussion on how these radii are computed refer to the methodology of \citet{2019ApJ...878..136Z} and \citet{2019ApJ...885..166Z}. 

\subsection{Cluster Membership}\label{sec: cluster membership}

We select cluster members from the APOKASC3 catalog for the open clusters NGC 6819 and NGC 6791 to conduct our analysis of the single-value scaling relations. We pick these clusters as they are well studied in the literature (e.g. \citet{2012ApJ...757..190C, 2011A&A...530A.100H, 2012MNRAS.419.2077M}) and sample interesting age and metallicity domains. As discussed in greater detail below, NGC 6819 is an intermediate-aged system, close to solar in abundance, and it therefore tests the scaling relations in stars that left the main sequence with convective cores. NGC 6791 is an old and very metal rich system, therefore it tests the scaling relations for stars that would have had radiative cores on the main sequence. \citep{2011ApJ...729L..10B, 2011A&A...525A...2B, 2018MNRAS.474.4810S}.  Additionally, while all clusters show some deviations from single-star evolution models, NGC 6819 has prior evidence for an unexpectedly wide mass range in the giants, while NGC 6791 also has a number of unusual evolved stars \citep{2017MNRAS.472..979H, 2018ApJS..239...32P, 2020Ap&SS.365...24G}.

We find few luminous giant stars with valid asteroseismic parameters and \textit{Gaia} radii in the APOKASC3 catalog. In NGC 6819, we find four luminous giant stars and in NGC 6791 we find one. We elect to remeasure the asteroseismic parameters from the power spectrum for the luminous giant in NGC 6791. The \numax parameter is determined using the method from \citet{2013A&A...559A.137M} which details how to measure \numax at low frequencies. \Dnu is measured using an auto-correlation function also described in \citet{2013A&A...559A.137M} with the universal red giant oscillation pattern given in \citet{2011A&A...525L...9M}. This value and the values for the luminous giants are given in  Table \ref{tab: NGC 6791 seismic measurements}. We use the MESA stellar evolution grid discussed in Section \ref{sec:model methods} to determine the \Fdnu correction factor for luminous giants without an \Fdnu value in NGC 6791 and NGC 6819. This table requires an input mass to compute \Dnu, so we use the median mass from the clusters as an input. With all of these components we are able to compute the asteroseismic masses for luminous giants in the clusters using the single parameter scaling relations.

\begin{table*}
\footnotesize
    \centering
    \begin{tabular}{c c c c c c c c}
         Cluster & KIC & \numax & $\sigma$ \numax & \Dnu & $\sigma$ \Dnu & \Fdnu & Remeasured Properties?  \\
                -  & - & [$\mu Hz$]   & [$\mu Hz$] & [$\mu Hz$] & [$\mu Hz$]  & - & Y/N\\
         \hline
        NGC 6819 & 5113061 &4.537 & 0.133 & 0.820 & 0.045 & 1.035 & N  \\

        NGC 6819 & 5112481 &5.359 & 0.141 & 0.882 & 0.029 & 1.028 & N  \\

        NGC 6819 & 5024851 &4.084 & 0.025 & 0.749 & 0.005 & 1.034 & Y\\

        NGC 6819 & 5024456 &4.241 & 0.171 & 0.717 & 0.046 & 1.031 & Y  \\

        NGC 6791 & 2571093 &2.041 & 0.065 & 0.471 & 0.031 & 1.041 & Y  \\

        \hline
    
    \end{tabular}

    \caption{Asteroseiemic measurements for luminous giants in NGC 6791 and NGC 6819.}
    \label{tab: NGC 6791 seismic measurements}
\end{table*}

\subsubsection{NGC 6819}

The open cluster NGC 6819 has been extensively studied due to its location in the \textit{Kepler} field. In \citet{2018ApJS..239...32P}, it was used as a fundamental mass calibrator for the APOKASC2 asteroseismic catalog. In that work they used eclipsing binary data from \citet{2016AJ....151...66B} to infer a red giant branch mass of $1.55 \pm 0.04 M_{\odot}$. \citet{2017MNRAS.472..979H} did an independent asteroseismic analysis which inferred a higher mass in both the RGB and RC of $1.61 \pm 0.02 M_{\odot}$ and $1.64 \pm 0.02$ respectively. These values are still in marginal statistical agreement with the eclipsing binary inference. We use two comparison points for the mass of the cluster. First we use the inferred \citet{2017MNRAS.472..979H} RGB mass. We also compare to an isochrone with properties discussed in Sec. \ref{sec: isochrone fit}.

NGC 6819 has also been a prime system for studying rotation age relationships for stars older than 1 Gyr, which has led to a renewed interest in the age of the cluster \citep{2015Natur.517..589M, 2023ApJ...947L...3B, 2013MNRAS.430.3472B}. The cluster has a measured turnoff age  of $\approx 2.25$ Gyr  \citep{2011ApJ...729L..10B, 2016AJ....151...66B}. This implies red giant masses in the domain consistent with both \citet{2016AJ....151...66B} and \citet{2017MNRAS.472..979H}. The cluster metallicity is well constrained to solar abundances by spectroscopic data (e.g. \citet{2015AJ....149..121L}).

We use \citet{2023A&A...673A.114H} to build an initial membership list. When we cross reference this list with APOKASC3, we reduce the sample from $2311$ stars to $53$ stars. We do a $3 \sigma$ cut using the \textit{Gaia} DR3 parallaxes and no stars are excluded. Using $3 \sigma$ cuts on metallicity and alpha abundance, we fine one $[\alpha/Fe]$ peculiar star, $KIC \ 5199859$, and omit it from the analysis sample. One star in the sample has an unknown evolutionary state classification and is therefore also excluded from the analysis. This brings the final sample used for analysis to 28 RGB stars and 22 red clump stars. Following the parallax and chemical cuts, the cluster's RGB has $[Fe/H] = 0.031 \pm 0.008$ and $[\alpha/Fe] = 0.006 \pm  0.002$. The red clump (RC) has $[Fe/H] = 0.055 \pm 0.005$ and $[\alpha/Fe] = -0.001 \pm 0.001$. 

In addition to the parallax and chemical cuts, we remove mass outliers to determine the final analysis sample for the clusters. We do a $3 \sigma$ mass cut for the RGB stars and a more generous $5 \sigma$ cut for the RC stars. We do this to reject stars that are not products of single star evolution, such as merger products. We allow a higher threshold for the RC mass cut to account for a larger expected spread in mass in the RC than on the RGB due to factors such as a dispersion in mass loss at the tip of the RGB. See 
\citet{2024arXiv241000102P} for further discussion. We remove stars that would be cataloged as mass outliers in either the APOKASC3 catalog mass, or the masses computed using the single parameter scaling relations given in Eqs. \ref{eq: mass scaling, numax only} and \ref{eq: mass scaling, dnu only}. If a star fails the sigma selection criteria in any of these mass categories, it is excluded from the analysis sample. 

\subsubsection{NGC 6791}

NGC 6791 has a very well characterized red giant branch mass because it has an exceptionally precise eclipsing binary mass measurement in the turnoff region \citep{2011A&A...525A...2B}.  The open cluster NGC 6791 was therefore also used as a fundamental mass calibrator in \citep{2018ApJS..239...32P}. In that work they used eclipsing binary data from the detailed studies by \citep{2011A&A...525A...2B} to infer a red giant branch mass of $1.15 \pm 0.02 M_{\odot}$.

The cluster metallicity is extremely high, especially for such an old system, and is well constrained by spectroscopic data \citep{2012A&A...541A..64J, 2018ApJ...867...34V}. Because of its old and metal rich nature, there have been detailed studies of its age and properties. The cluster turnoff age of $\approx 8 \ Myr$ also implies red giant masses in the domain consistent with the eclipsing binaries \citep{2011A&A...525A...2B}. 

\begin{figure}
    \centering
    \includegraphics[width=0.5\textwidth]{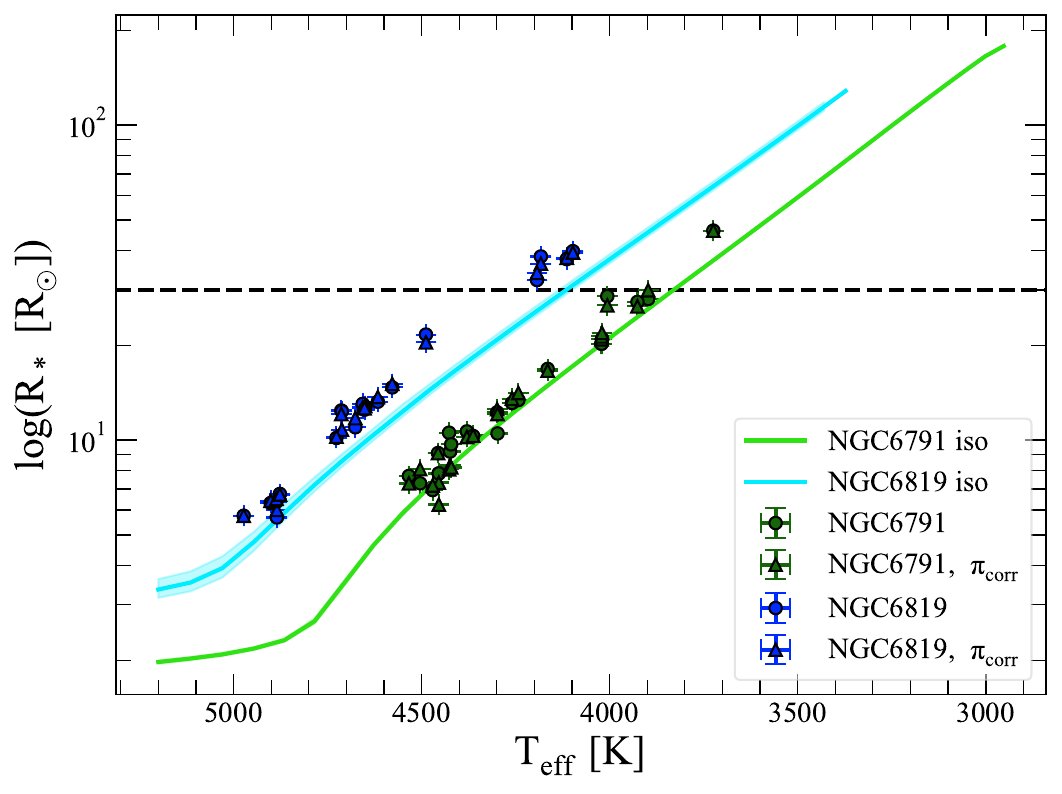}
    \caption{Radius - \Teff diagram for NGC 6791 and NGC 6819. Lines indicate best fit isochrone and shaded regions indicate errors on best fit isochrone. The triangular points are the parallactic radius before correcting random errors using the cluster parallaxes. The circular points are the parallactic radii after correcting for random errors using the cluster parallax. See Sec. \ref{sec: isochrone fit} for the parameters of the isochrones}
    \label{fig: cluster HRD}
\end{figure}

We again start with \citet{2023A&A...673A.114H} to build a membership list. After cross-referencing with APOKASC3, we have $60$ stars. We find $2 \ [\alpha/Fe]$ peculiar RGB stars. Following parallax and chemical cuts, our NGC 6791 sampe is left with 55 stars, 30 of which are giants, 25 of which are clump stars. The final abundances of the cluster's RGB are $[Fe/H] = 0.294 \pm 0.029$ and $[\alpha/Fe] = 0.030 \pm 0.009$. The RC has $[Fe/H] = 0.320 \pm 0.008$ and $[\alpha/Fe] = 0.033 \pm 0.003$. A radius - \Teff diagram for stars in the NGC 6791 and NGC 6819 is provided in Fig \ref{fig: cluster HRD}

\subsubsection{Cluster isochronal masses} \label{sec: isochrone fit}
The APOKASC2 catalog used eclipsing binaries near the cluster turnoff to infer predicted masses of stars on the clusters giant branches. We complement this analysis by adopting literature age estimates to predict masses. Here, we use isochrones from the MIST stellar evolution database to determine the cluster isochronal mass \citep{2016ApJS..222....8D, 2016ApJ...823..102C, 2011ApJS..192....3P, 2013ApJS..208....4P, 2015ApJS..220...15P, 2018ApJS..234...34P}. For NGC 6819, we adopt an age of $2.21 \pm 0.3 Gyr$ as measured by \citet{2016AJ....151...66B} and a metallicity equal to the average metallicity of the cluster. This isochrone has a mass of $1.59 \pm 0.07 \ M_{\odot}$. For NGC 6791, we use an isochrone with an age of $8.3 \pm 0.3 \ Gyr$ as measured by \citet{2021A&A...649A.178B} and a metallicity equal to the cluster's mean metallicity. However, as noted in (Joyce, Silva Aguirre) papers, there are age systematics at the $5\%$ or higher level even for stars without convective cores or mass loss, corresponding to a systematic uncertainty of order $0.4$ Gyr for this system \citep{2018MNRAS.475.5487S, 2023ApJ...946...28J}. We therefore adopt an isochrone mass of $1.14 \pm 0.01 M_{\odot}$. Figure \ref{fig: cluster HRD} shows the radius-\Teff diagram for the clusters and their isochrones.

\begin{table}
\footnotesize
    \centering
    \begin{tabular}{c c c c}
         Cluster & Age & $[Fe/H]$ & $M_{iso}$\\
                -  & Gyr & $[Fe/H]_{\odot}$ & $M_{\odot}$\\
         \hline
        NGC 6819 & $2.2 \pm 0.3$ & 0.029 & $1.59 \pm 0.07$\\
        
        NGC 6791 & $8.3 \pm 0.3$ & 0.298 & $1.15 \pm 0.01$\\

        \hline
    
    \end{tabular}

    \caption{Adopted isochrone properties for open clusters NGC 6819 and NGC 6791}
    \label{tab: cluster iso}
\end{table}

\subsubsection{High-$\alpha$ Stars}

\begin{figure}
    \centering
    \includegraphics[width=0.5\textwidth]{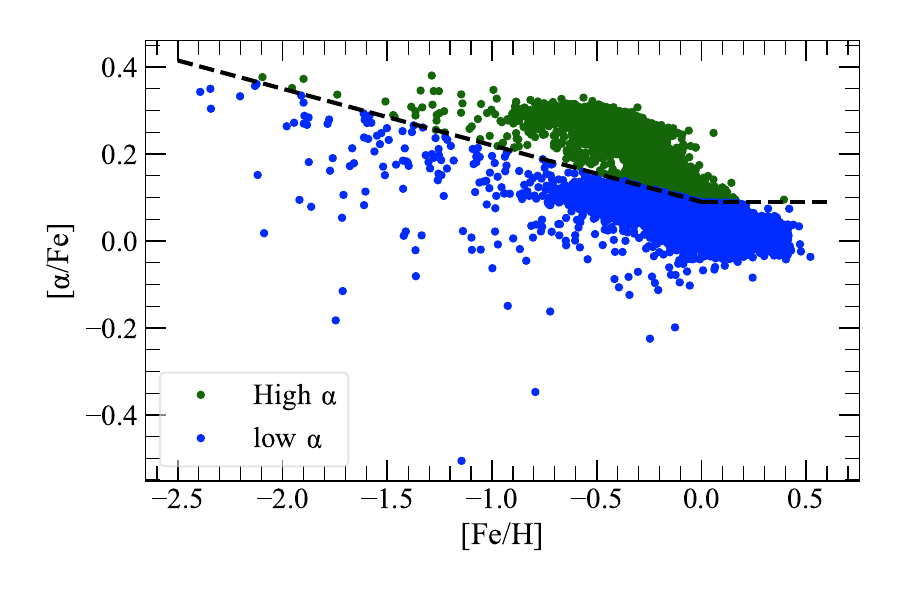}

    \caption{High-$\alpha$ sequence selection. High-$\alpha$ stars are green points above the dotted line. The dotted line is given by the prescription in \citet{2024MNRAS.530..149R}}
    \label{fig: high alp selction}
\end{figure}

In addition to the cluster analysis, we use the high-$\alpha$ sequence as a "pseudo" cluster in our analysis of the single parameter asteroseismic scaling relations. \citet{2021A&A...645A..85M} identified the high-$\alpha$ sequence as the thick disk of the Milky Way with a well defined mean age of $\approx 11$ Gyrs old and a modelling dependent dispersion in age of $\approx 1$ Gyr. In the APOKASC3 sample, the mean age of the high-alpha RGB sample is $9.14 \ Gyr$ with a $+/- 0.05 \ Gyr$ random error and a $\pm 0.9 \ Gyr$ standard error. This age is about $2.2 \sigma$ smaller than the \citet{2021A&A...645A..85M} value. In both studies however, the dispersion in the age of the high-$\alpha$ sequence is small. There is an interesting sub-population of more massive $\alpha$ rich stars, which could be a mixture of truly young stars and merger products, but the large majority of the stars are truly low mass and old \citep{2015A&A...576L..12C, 2020ApJ...903...12S, 2021A&A...645A..85M, 2021ApJ...922..145Z}. It is therefore reasonable to use the inferred masses as a function of radius to test for radius-dependent deviations from the scaling relations. To select the high-$\alpha$ stars, we adopt the definition for the high-$\alpha$ and low-$\alpha$ sequences given in \citet{2024MNRAS.530..149R}. We also perform a $5 \sigma$ mass cut on the APOKASC3 catalog mass and the single-parameter scaling relation masses to remove young, over-massive high - $\alpha$ stars as noted in \citet{2021A&A...645A..85M} and \citet{2016MNRAS.456.3655M}. Additionally, to reduce the impact of these stars on our sample, we use median statistics.  Figure \ref{fig: high alp selction} shows our sample of $1552$ high - $\alpha$ stars in APOKASC3. The median mass of stars on the high-$\alpha$ sequence with radii less than $30 R_{\odot}$ is $1.04$ with a characteristic dispersion given by the median absolute deviation over the square root of the number of stars of $0.002 \ M_{\odot}$. We compute \Fdnu correction factors, masses and radii for stars with \numax values less than $1 \mu Hz$ as these values are not computed in the APOKASC3 catalog. We set the $F_{\nu_{max}} = 1$ for all stars in our high-$\alpha$ sample.

\section{Cluster Analysis}\label{sec: cluster methods}
In this work, we are interested in how each of the asteroseismic parameters, \Dnu and \numax, individually affect our ability to determine fundamental stellar properties using the scaling relations. To do this, we isolate the two terms and create scaling relations that are dependent on only \numax and \Dnu respectively. These relationships are defined as: 

\begin{equation}\label{eq: mass scaling, numax only}
    \frac{M_*}{M_{\odot}} = \bigg(\frac{\nu_{max}}{\nu_{max, \odot}}\bigg) \bigg(\frac{R_{Gaia}}{R_{\odot}}\bigg)^{2} \bigg(\frac{\text{T}_{\text{eff}}}{\text{T}_{\text{eff}, \odot}}\bigg)^{1/2}
\end{equation}

for the \numax only scaling relation, and 

\begin{equation}\label{eq: mass scaling, dnu only}
    \frac{M_*}{M_{\odot}} = \bigg(\frac{F_{\Delta \nu}\Delta \nu}{\Delta \nu_{\odot}}\bigg)^2 \bigg(\frac{R_{Gaia}}{R_{\odot}}\bigg)^{3} 
\end{equation}

for the \Dnu only scaling relation. The APOKASC3 catalog contains an additional correction term, \Fnmax, that is used to correct \numax in the scaling relations. However, as we are interested in tracing the origin of deviations from scalings in luminous stars, we test the seismic masses computed with and without this correction term. However, rather than adopting the catalog \Fnmax values which vary as a function of \numax along the giant branch, we adopt a constant \Fnmax of $0.9959$ in RGB stars and $0.9937$ in RC stars. We use this constant as we are interested in trends in the deviations rather than individual star zero point shifts. 

We use Eqns. \ref{eq: mass scaling, numax only} and \ref{eq: mass scaling, dnu only} to determine the extent to which the \numax and \Dnu parameters are responsible for the observed breakdown in the scaling relations. We apply the single scaling relations to stars in the open clusters NGC 6791 and NGC 6819 to compute the masses of giant stars. Our approach is similar to the \citet{2012MNRAS.419.2077M} that looked for mass loss signatures between red clump and giant branch stars in NGC 6819 and NGC 6791. We use cluster data in this component analysis as clusters provide seismic-independent stellar radii and stellar masses - the parallactic radii and the cluster isochronal masses respectively. 

The errors on the single scaling relations are found via error propagation. For the \numax-only relation we have: 

\begin{equation}
    \frac{\sigma M}{M} = \sqrt{\bigg(\frac{\sigma \nu_{max}}{\nu_{max}}\bigg)^2 + \bigg(\frac{1}{2}\frac{\sigma T_{eff}}{T_{eff}}\bigg)^2 +  \bigg(2\frac{\sigma R_{Gaia}}{R_{Gaia}}\bigg)^2} \,
\end{equation}

and 

\begin{equation}
    \frac{\sigma M}{M} = \sqrt{\bigg(2\frac{\sigma \Delta \nu}{\Delta \nu}\bigg)^2 + \bigg(2\frac{\sigma F_{\Delta \nu}}{_{\Delta \nu}}\bigg)^2 +  \bigg(3\frac{\sigma R_{Gaia}}{R_{Gaia}}\bigg)^2} 
\end{equation}

for the \Dnu-only masses

\subsection{Parallactic Radius Errors}\label{sec:radius errors}

An additional advantage of using cluster data is that we are able to re-compute the parallactic radius of cluster members. This "corrected" radius reduces the random error in the parallactic radius and is given by: 

\begin{equation}\label{Eq: parallax corr}
    R_{Gaia, corr} = R_{Gaia} * (\frac{\pi_{*}}{\pi_{cluster}}), 
\end{equation}

where $\pi_{cluster}$ is the average parallax of the cluster. By adopting a single parallax for the cluster, we elect to ignore depth effects and differential reddenning. However, the random error in the parallax is dominated by the \textit{Gaia} zero point, such that these effects will be small. Results presented below use Eq. \ref{Eq: parallax corr} as the seismic-independent radius to compute the mass in the single-parameter scaling relations.


\subsection{Single Scaling Relation Results}

In this section, we detail how the \numax and \Dnu scaling relation masses perform relative to seismic-independent masses. 

\subsubsection{Clusters}\label{sec: cluster results}

\begin{figure*}
    \centering
    \includegraphics[width=\textwidth]{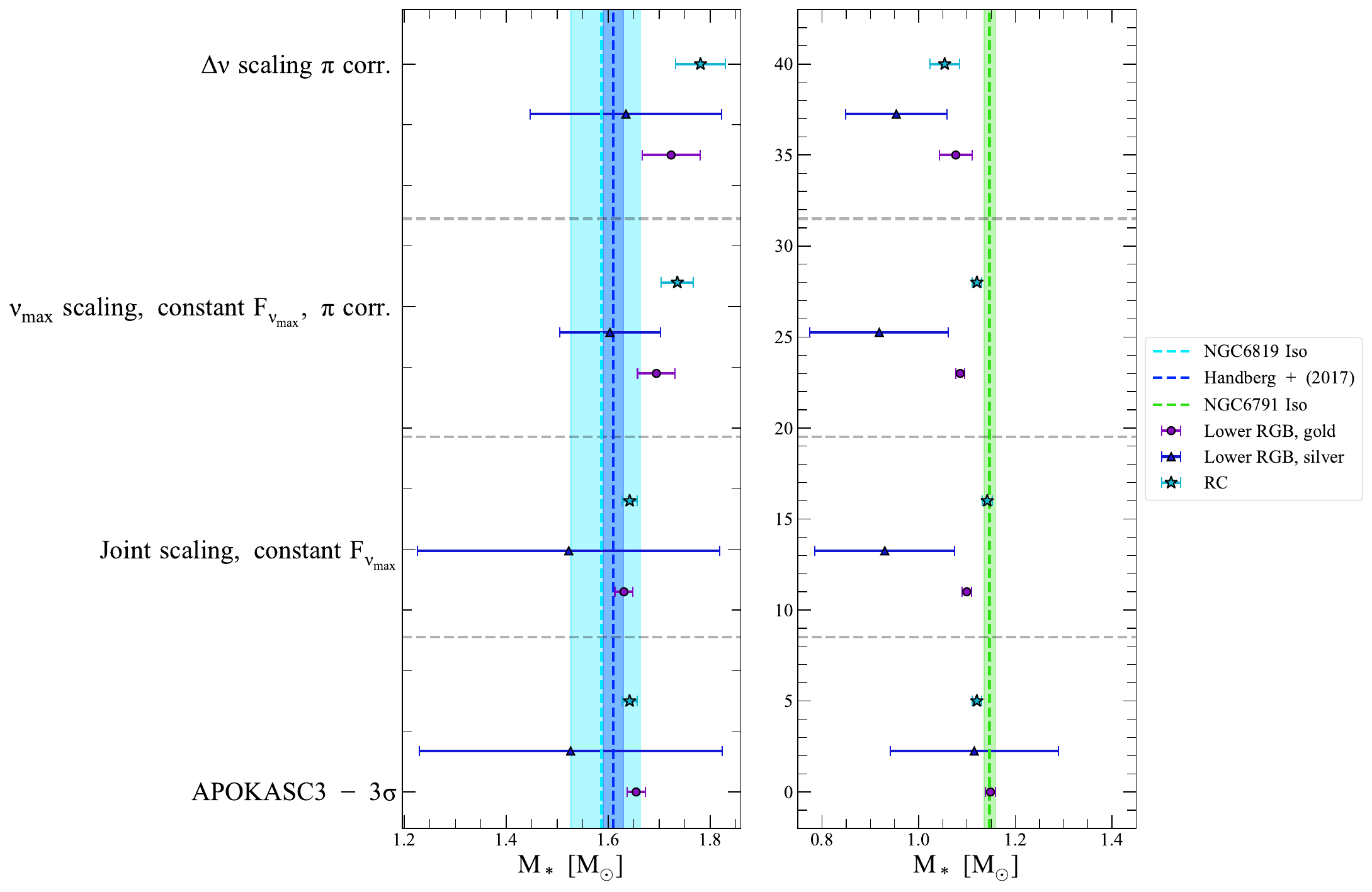}
    \caption{Cluster weighted average mass results. Left panel: NGC 6819 mass results compared to an isochrone mass and the mass of NGC 6819 given in \citet{2017MNRAS.472..979H}. Right: NGC 6791 mass results compared to an isochrone mass. The parameters of the isochrone can be found in section \ref{sec: isochrone fit}. This plot shows results for the lower RGB gold sample, the lower RGB silver sample and the RC. The luminous giants are omitted due to small number statistics. The error bars on the weighted mean masses are the standard errors of the weighted mean.}
    \label{fig: cluster summary results}
\end{figure*}

\begin{figure*}
    \centering
    \includegraphics[width=\textwidth]{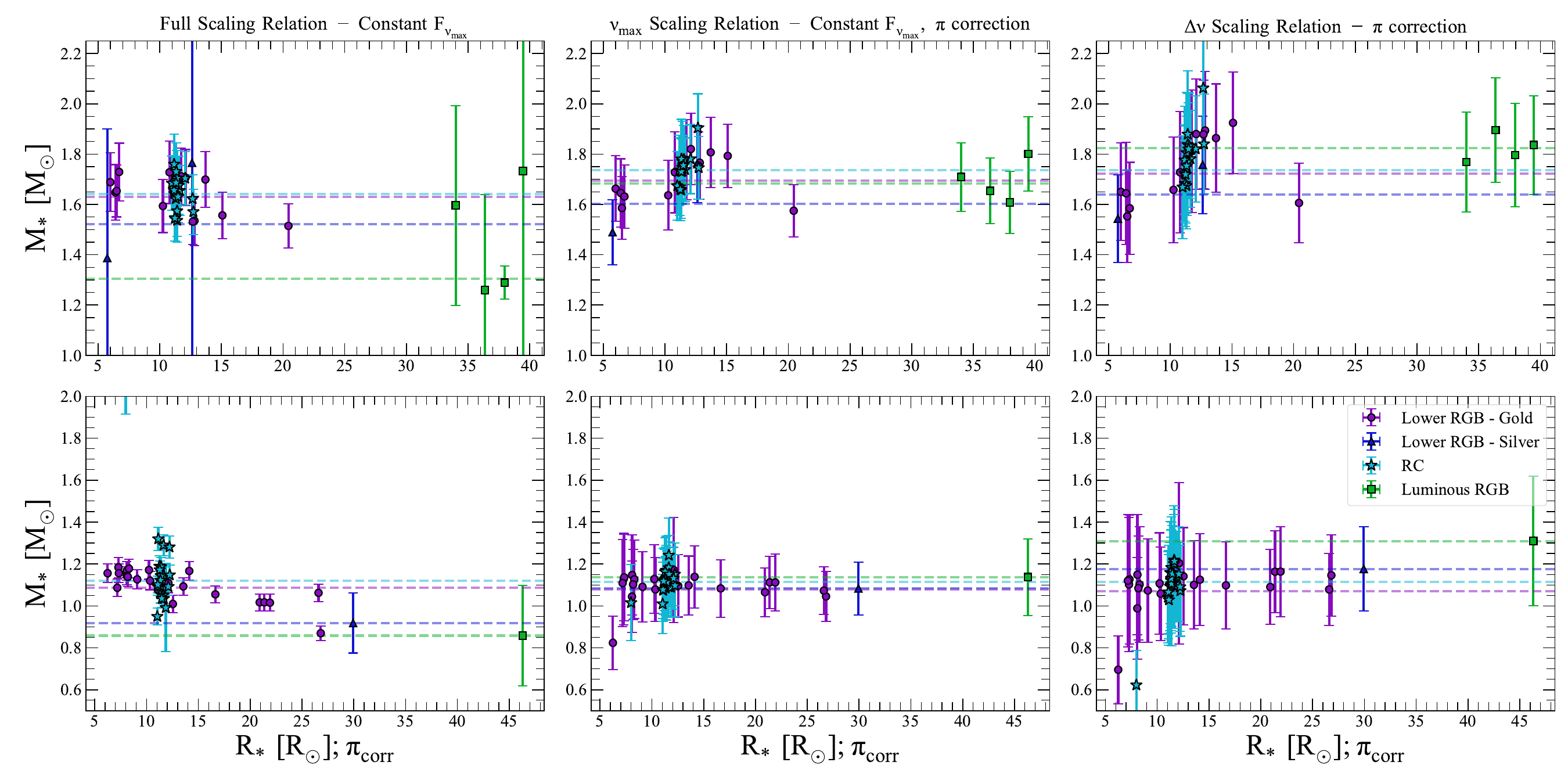}
    \caption{Component analysis of the scaling relations using clusters in $M_*  \ v. \ R_*$ plane. Different colors indicate different samples within the cluster. The colored dotted lines are the weighted mean masses of a particular sample within a cluster. Left: Masses computed using the full scaling relation with a constant \Fnmax value for RGB and RC stars. Middle: Masses using the \numax only scaling relation with a constant \numax for RGB and RC stars and parallax-corrected Gaia radii. Right: Masses using the \Dnu only scaling relation with parallax-corrected Gaia radii. See Sec. \ref{sec: cluster results} for discussion about this figure.}
    \label{fig: cluster components}
\end{figure*}

The summary results for multiple test cases of the joint and single-scaling relations in NGC 6791 and NGC 6819 is shown in Fig. \ref{fig: cluster summary results}. Tables \ref{table:cluster results NGC 6819} and \ref{table:cluster results NGC 6791} provide the weighted mean average masses in different stellar samples in the clusters. These tables provide masses for the joint scaling relation with and without an \Fnmax term, the \numax scaling relation with and without a parallax corrected \textit{Gaia} radius, and the \Dnu scaling relation with and without a parallax corrected \textit{Gaia} radius.  We also would add a note of caution that the method used to compute the seismic masses has an impact on the weighted average mass computed for the sample. Factors such as computing the seismic masses in inverse space, or outlier selection do have an impact on the final calculation. We measure the weighted mean masses and the standard error on the weignted mean masses for different groups of stars within the cluster. These groups, from the APOKASC3 definitions, are lower RGB gold stars, lower RGB silver stars, RC stars and luminous RGB stars. The lower RGB gold stars are stars with \textit{Gaia} radii $\leq \ 30 R_{\odot}$, measured \Dnu and \numax values, and the most precise data. The lower RGB silver sample also has  \textit{Gaia} radii $\leq \ 30 R_{\odot}$ and measured \Dnu and \numax values, but the data is less precise in these stars. An example of one of our test cases is shown for cluster members in Fig. \ref{fig: cluster components}. 

Shown in Fig. \ref{fig: cluster summary results}, we find in agreement with prior results, that the mean masses of NGC 6819 RC and RGB stars are higher than predicted, but consistent within the errors. Our lower RGB gold sample mass estimate for the joint scaling relationship, $1.63 \pm 0.02 M_{\odot}$, is $1.8$ sigma higher than the \citet{2018ApJS..239...32P} value and $0.65$ sigma higher than our isochrone fit.  We are also in excellent, $0.78$ sigma agreement with the \citet{2017MNRAS.472..979H} value of $1.61 \pm 0.02 M_{\odot}$. Despite being a seismic-dependent analysis, we elect to include the \citet{2017MNRAS.472..979H} cluster mass as a comparison point as it represents an entirely different approach to the seismic data. Specifically, this study measures seismic masses through an analysis of the individual modes in the frequency spectra rather than using global properties. For the single scaling relations, if we adopt the mean cluster parallax, we get a RGB mass estimate of $1.72 \pm 0.06 M_{\odot}$ for the \Dnu relation, which is in statistical agreement with the joint scaling result, but higher than predicted. The \numax relation gives a somewhat lower value at $1.69 \pm 0.04 M_{\odot}$. The higher than expected asteroseismic masses compared to the eclipsing binary fit and the isochrone fit suggest that the cluster age is moderately younger than these metrics predict. Using the MIST isochrones and the asteroseismic mass, we determine the cluster is $2.1 \pm 0.24$ Gyrs old. Unlike the case of NGC 6791, the eclipsing binary is complex to analyze, and systematic errors in the masses of the level required to explain the discrepancy are reasonable. 

This is also true for the lower RGB silver sample though the lower RGB silver sample mean has a larger uncertainty. The RC sample also tends to be more massive than isochronal predictions, with this effect being larger in the single-parameter scaling relations. Again, these results are within $3 \sigma$ of the isochrone fit, so are not statistically significant. 

NGC 6791 displays a similar level of statistical agreement between the seismic masses inferred by the different scaling relations and the isochronal mass. Following $3 \sigma$ mass cuts, the lower RGB gold stars have APOKASC3 catalog masses that are remarkably consistent with the isochrone mass - less than $1$ sigma away. The uncalibrated joint scaling relation mass of $1.08 \pm 0.01$ is statistically inconsistent with the isochrone mass by $4.1$ sigma due to the small uncertainty on the mean measurement. This inconsistency can be traced to the adopted \Fnmax parameter. If we adopt the catalog \Fnmax values - which vary as a function of radius across the RGB - we find the lower RGB gold sample has a mass of $1.15 \pm 0.01 M_{\odot}$ which are in statistical agreement with the isochrone. Many of the RGB stars in NGC 6791 exist in the radius domain where they have a larger \Fnmax value than the constant \Fnmax value used in this analysis. As we are interested in understanding the origin of this radius-dependent offset, we hold this value fixed for our analysis. For the single-parameter mean parallax relations, we find that both the \Dnu-only relation and the \numax-only relation is in statistical agreement with the isochrone. 

Due to a a small number of available luminous giant stars in the clusters, we do not show results for the standard mean masses of luminous giants in the cluster in Fig. \ref{fig: cluster summary results}. In NGC 6791, we have one luminous giant star in our analysis and this star shows substantial deviations from the isochrone mass in the joint and \Dnu-only scaling relations. However, the uncertainty on this star's mass is high such that the mass difference is not statistically significant. In NGC 6819, we have four luminous giant stars. The luminous giant joint scaling relation mass is $3.35$ sigma away from the isochrone mass. While the robustness of our conclusions about luminous giant stars in these open clusters is limited by the small number of available stars, we show that stars with similarly large uncertainties, the lower RGB silver sample, are consistent with isochronal mass. Therefore, if the errors on these stars are trusted, we demonstrate there is a statistically significant deviation in the luminous giant stars of NGC 6819. This fortifies our claim that we cannot reliably predict luminous giant masses. 
    
Some studies have found evidence for mass dispersion in the RGB of NGC 6819 \citep{2017MNRAS.472..979H, 2018ApJS..239...32P}. However, following $3 \sigma$ mass cuts to remove likely merger products, we find no evidence of an anomalous dispersion in the cluster's mass. The standard deviation in mass on the lower RGB gold sample is $0.097 \ M_{\odot}$  and the predicted standard deviation given by the individual errors on the masses is $0.109 \ M_{\odot}$, which is in reasonable agreement. Similarly, NGC 6791 has a standard deviation in the lower RGB gold sample mass dispersion of $0.049 \ M_{\odot}$ and a predicted dispersion of $0.052 \ M_{\odot}$ for the cluster dispersion. Therefore, the dispersion in mass in NGC 6819 on the lower giant branch is reasonable given both the predicted dispersion from the mass errors and the similar agreement between the predicted and observed dispersion in NGC 6791. 

While not the main focus of this study, previous studies of NGC 6819 have found anomalous mass members in the cluster \citet{2017MNRAS.472..979H}. We check our sample to see if these members are present. Five over-massive members, \textit{KIC 5024414}, \textit{KIC 5024476}, \textit{KIC 5023953}, \textit{KIC 5112880}, and \textit{KIC 5024272} are in our initial membership list, but cut from the sample by mass outlier rejection. One of the over-massive members identified in \citet{2017MNRAS.472..979H}, \textit{KIC 5112361}, is not in the membership list we use, indicating with updated \textit{Gaia} DR3 parallax information is not associated with the cluster. Only one under-massive member identified in \citet{2017MNRAS.472..979H} is in our sample, \textit{KIC 5113061}, which is a luminous giant and therefore not subject to the same mass outlier rejection criteria as the lower RGB and RC stars. \textit{KIC 4937011} is an under-massive, Li-rich RC star and is cut during mass outlier rejection along with \textit{KIC 4937770}. Our membership list is from \citet{2023A&A...673A.114H} which uses Gaia DR3 data. This indicates that many of the over-massive members present in the \citet{2017MNRAS.472..979H} study are cluster members given updated parallax and kinematic information from \textit{Gaia DR3}. Over-massive members are not entirely unexpected in open clusters. For example, blue stragglers, a byproduct of non-single star evolution, produce stars that appear anomalously massive. 

Another interesting facet of the cluster analysis is our RC results. In NGC 6819, we find that the RC masses are larger than the lower RGB gold sample masses. This is expected from prior results \citep{2017MNRAS.472..979H}. We emphasize however that the RC masses are always larger than the RGB masses across the different scaling relation tests. In NGC 6791, we find that there is little mass loss between the RGB and the RC, which is consistent with the literature \citep{2012MNRAS.419.2077M, 2018ApJS..239...32P}.

\begin{table*}
\footnotesize
    \centering
    \begin{tabular}{c c c c c}
        Sample & Method & \Fnmax used & Mass & $\sigma$ Mass \\
                 - & - & - & [$M_{\odot}$] & [$M_{\odot}$] \\
         \hline
                Lower RGB - gold & APOKASC3 Mass cut & - &  1.655 & 0.018 \\
                Lower RGB - silver & APOKASC3 Mass cut & - &  1.526 & 0.297 \\
                Red Clump & APOKASC3 Mass cut & - &  1.642 & 0.015 \\
                Luminous RGB & APOKASC3 Mass cut & - &  1.585 & 0.061 \\
                Lower RGB - gold & Joint scaling & constant \Fnmax & 1.631 & 0.018 \\
                Lower RGB - silver & Joint scaling & constant \Fnmax & 1.522 & 0.296 \\
                Red Clump & Joint scaling & constant \Fnmax &  1.642 & 0.015 \\
                Luminous RGB & Joint scaling & constant \Fnmax &  1.304 & 0.049 \\
                Lower RGB - gold & \numax scaling & constant \Fnmax & 1.694 & 0.037 \\
                Lower RGB - silver & \numax scaling & constant \Fnmax &  1.603 & 0.099 \\
                Red Clump & \numax scaling & constant \Fnmax &  1.735 & 0.031 \\
                Luminous RGB & \numax scaling & constant \Fnmax &  1.684 & 0.067 \\
                Lower RGB - gold & \Dnu scaling & - &  1.723 & 0.056 \\
                Lower RGB - silver & \Dnu scaling & - &  1.639 & 0.130 \\
                Red Clump & \Dnu scaling & - &  1.781 & 0.048 \\
                Luminous RGB & \Dnu scaling & - &  1.823 & 0.101 \\
                
         \hline 
    
    \end{tabular}
    
    \caption{Cluster mass results for NGC 6819.}
    \label{table:cluster results NGC 6819}
\end{table*}

\begin{table*}
\footnotesize
    \centering
    \begin{tabular}{c c c c c}
        Sample & Method & \Fnmax used  & Mass & $\sigma$ Mass \\
             - & -  & -  & [$M_{\odot}$] & [$M_{\odot}$] \\
         \hline
            Lower RGB - gold & APOKASC3 Mass cut & - & 1.148 & 0.010 \\
            Lower RGB - silver & APOKASC3 Mass cut & - & 1.115 & 0.174 \\
            Red Clump & APOKASC3 Mass cut & - & 1.120 & 0.010 \\
            Luminous RGB & APOKASC3 Mass cut & - & 0.930 & 0.258 \\
            Lower RGB - gold & Joint scaling & constant \Fnmax & 1.086 & 0.009 \\
            Lower RGB - silver & Joint scaling & constant \Fnmax & 0.918 & 0.143 \\
            Red Clump & Joint scaling & constant \Fnmax & 1.120 & 0.010 \\
            Luminous RGB & Joint scaling & constant \Fnmax & 0.858 & 0.239 \\
            Lower RGB - gold & \numax scaling & constant \Fnmax & 1.078 & 0.033 \\
            Lower RGB - silver & \numax scaling & constant \Fnmax &  1.083 & 0.126 \\
            Red Clump & \numax scaling & constant \Fnmax & 1.115 & 0.033 \\
            Luminous RGB & \numax scaling & constant \Fnmax & 1.137 & 0.183 \\
            Lower RGB - gold & \Dnu scaling & - & 1.070 & 0.049 \\
            Lower RGB - silver & \Dnu scaling & - & 1.176 & 0.201 \\
            Red Clump & \Dnu scaling & - & 1.076 & 0.048 \\
            Luminous RGB & \Dnu scaling & - & 1.309 & 0.309 \\
            
        \hline 
    
    \end{tabular}
    
    \caption{Cluster mass results for NGC 6791.}
    \label{table:cluster results NGC 6791}
\end{table*}

\subsubsection{High-$\alpha$ sequence}\label{sec: high-alpha results}

\begin{figure*}
    \centering
    \includegraphics[width=\textwidth]{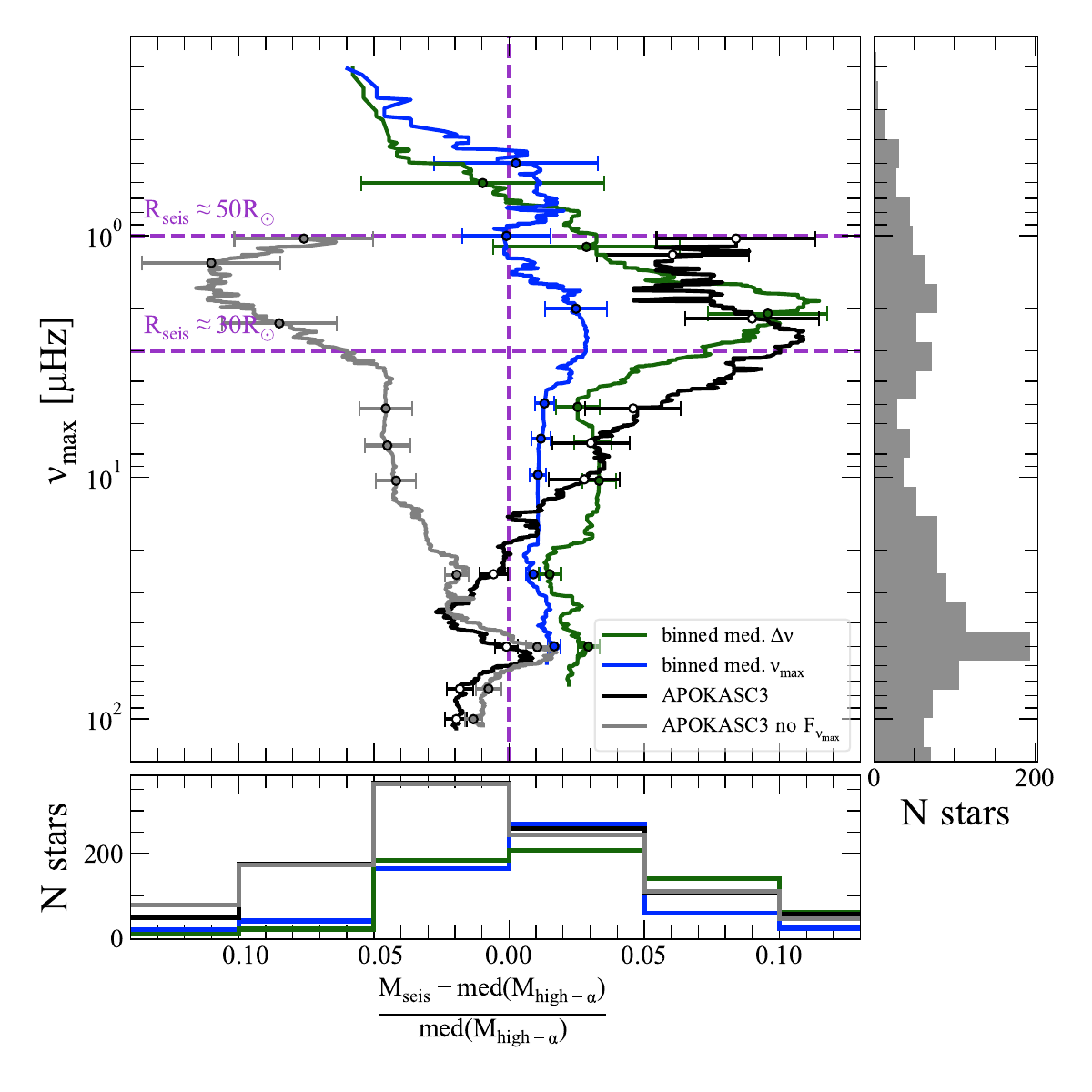}
    \caption{Component analysis of the scaling relations using high-$\alpha$ sequence stars. The x-axis is the fractional mass difference between mass computed using the scaling relations and the median mass of the high-$\alpha$ sequence. The purple vertical dashed line is the zero-offset point and the purple horizontal dashed line is the \numax value where we consider stars to be luminous giants. The green and blue lines are the rolling medians of masses computed using only \Dnu and \numax respectively. The window size to compute the rolling median is $150$ points. The black line is the full scaling relation mass using the \Fnmax correction parameter. The grey line is the full scaling relation mass, omitting the \Fnmax correction factor. Points are the median values of the offset at specific \numax values. The error bars on these points are the 1.48 *  MAD divided by the square root of the number of points used to compute the rolling median at a given point, which is set equal to the window size. The histogram on the y-axis shows the number of stars in \numax bins. The histogram on the x-axis shows the number of stars in bins of fractional difference between the seismic masses and the median high-$\alpha$ sequence mass. Section \ref{sec: high-alpha results} interprets the results of this figure.}
    \label{fig: High alp results}
\end{figure*}

\begin{figure}
    \centering
    \includegraphics[width=0.5\textwidth]{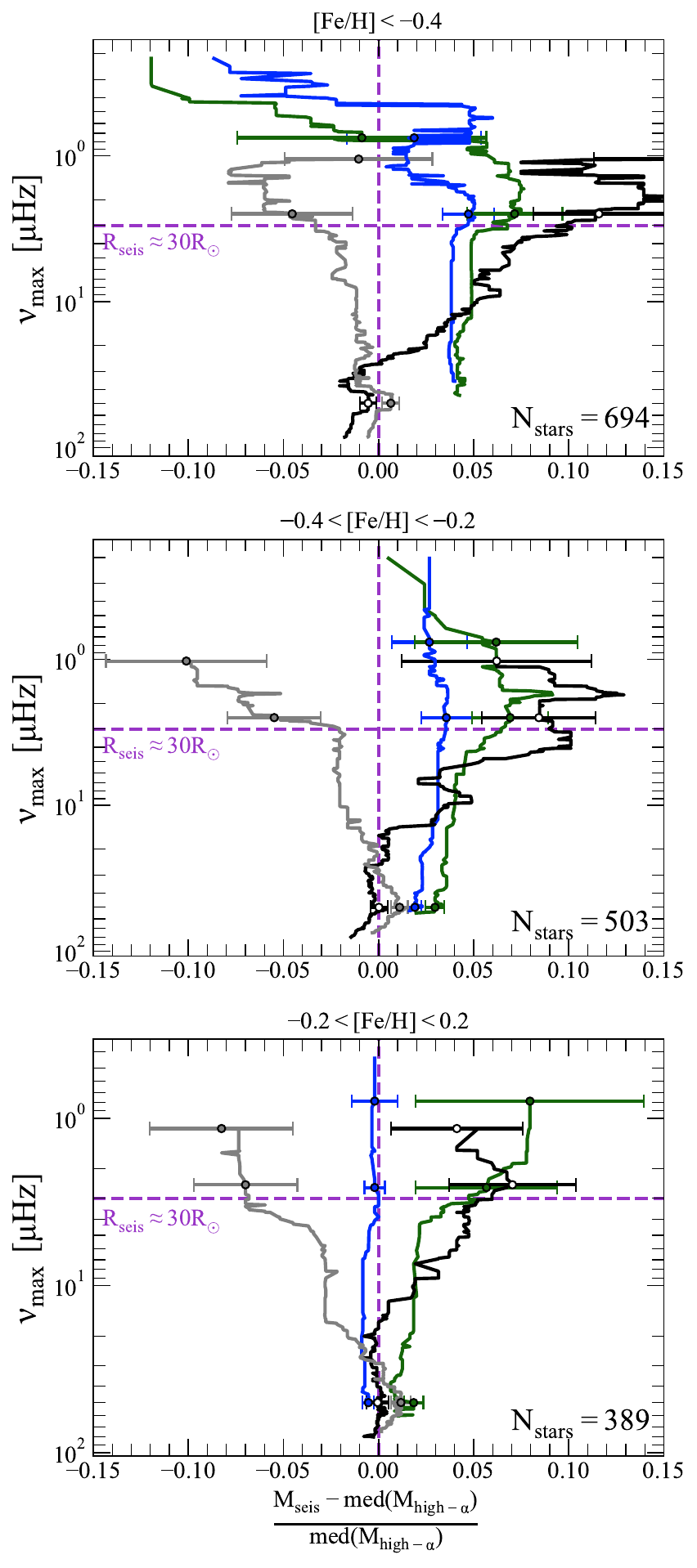}
    \caption{Component analysis of the scaling relations using high-$\alpha$ sequence stars binned by metallicity. The lines and axes of this figure follow the same convention as Fig. \ref{fig: High alp results}. The median mass of the high-$\alpha$ sequence, used for the reference point, is remeasured for each metallicity bin. From the top panel to the bottom panel is increasing metallicity. See Sec. 
    \ref{sec: high-alpha results} for details about this figure. }
    \label{fig: High alp results metallicity bins}
\end{figure}

Using the high-$\alpha$ sequence as a single mass population, we can test the single-scaling relation masses and the calibration of radii in the APOKASC3 system. Figure \ref{fig: High alp results} shows our results for the full high-$\alpha$ sample. We demonstrate that caution should be taken when using the single-parameter scaling relations - particularly the \Dnu-only relation in the luminous giant regime. We also show that using \Fnmax to calibrate in radius does not necessary calibrate the mass scale. Without a calibration, we clearly get masses in the \numax of $1$ to $10 \ \mu Hz$ that are too small - reinforcing the need for a calibration. However, when a calibration is included, our masses are higher than those at lower \numax, indicating that a radius calibration can leave a residual, if smaller, systematic in masses. 
\citet{2024arXiv241000102P} noted that ages for more luminous high-$\alpha$ stars were systematically lower than those for the lower RGB.  AGB stars are present on the upper RGB, which would lead to an over-estimate of ages and an underestimate of masses; note, however, that the radius calibration should be valid for both groups.

We find that the \numax-only relation produces masses that are consistent with the median mass of the high-alpha sequence. Using the median absolute deviation (MAD) of the points within a bin set by the window size of the rolling median calculation, we find across the RGB that the \numax-only relation is within 3 sigma of the median mass of the high-$\alpha$ sequence. The \Dnu-only relation produces masses that are more discrepant with the median mass of the high-$\alpha$ sequence. In the regime between $30$ and $50 \ R_{\odot}$, we find the greatest significance in this discrepancy. This indicates that in this regime issues in \Dnu are driving the observed breakdown in the seismic radii of luminous giants. At larger radii, the large random errors in our measurements are too large to say for certain if \Dnu is driving the discrepancy in this domain.

While both the \Dnu-only and \numax-only relation are in statistical agreement with the high-$\alpha$ sequence mass, the curvature of the single-scaling relations reveals insights into the breakdown of the scaling relations. The \numax-relation produces masses that are relatively constant moving up the RGB. The \Dnu-only relation however, produces over-massive stars moving towards lower \numax values, then begins to move towards under-massive stars in the most luminous giants. This curvature mirrors the curvature we see in the full scaling relationship. This indicates that issues in the joint scaling relation in luminous giants may be traced to \Dnu. 

In addition to determining the median deviations of the single-parameter scaling relations, we also determine whether we can distinguish between the two relationships. Using a Kolmogorov-Smirnov (KS) tests we confirm that the masses computed with the different single-parameter scaling relations have a statistically significant difference. 
The lower RGB, the luminous RGB, and in the ultra-luminous RGB samples all show statistically distinguishable distributions via the KS test.

The high-$\alpha$ sequence exists across a range of metallicities. Therefore, there is some intrinsic scatter in the mass of the high-$\alpha$ sequence. To isolate the intrinsic scatter of the high-$\alpha$ sequence due to metallicity from deviations due to the scaling relations we divide Fig. \ref{fig: High alp results} into metallicity bins. Three metallicity bins are shown in Fig. \ref{fig: High alp results metallicity bins}. By binning the high-$\alpha$ stars by metallicity, we do see less scatter in the masses due to reduced intrinsic scatter from metallicity changes. However, we still see clear deviations from the median mass of the high-$\alpha$ stars from the scaling relations at large radii. 

The behavior of the single-parameter scaling relations in the metallicity bins is similar to their behavior of the full high-$\alpha$ sample. In general, we see that the \numax-only relation produces results in agreement with the median mass of the high-$\alpha$ sequence in each metallicity bin. The \Dnu-only relation on the other hand produces masses that are more discrepant with the median mass of the high-$\alpha$ sequence and mirrors the behavior of the full APOKASC3 sample. The results from the high-$\alpha$ sequence analysis may point to the \Dnu relationship producing the most discrepant masses across all metallicity bins. However, this result is not at a 3 $\sigma$ level of significance. In combination with Fig. \ref{fig: High alp results}, our results in the different metallicity bins may allude to greater deviations in the \Dnu-only scaling relation and therefore underlying issues with \Dnu in the luminous giant regime. We also note that at the lowest metallicities, we see the largest deviations from the median mass of the high-$\alpha$ sequence across the entire giant branch. 

When paired with the cluster results, we recommend caution when using the single-parameter scaling relations in the luminous giant regime. There are indications that the \Dnu-only scaling relation contributes more greatly to the breakdown in the full scaling relations. This has implications on future work that relies on only one of the seismic parameters. In particular, the upcoming Nancy Grace Roman Space telescope will observe stars in the galactic bulge and determine masses and radii via asteroseismology. The asteroseismic masses and radii are planned to be determined via \numax only - which we have demonstrated are a reliable figure of merit on the lower RGB, but show some level of discrepancy in luminous stars. Additional data is needed to further investigate this result. 

We have a theoretical framework to compute a correction factor for \Dnu. Based on this and the apparent issues with the \Dnu - only scaling relation, we continue on to test how changing the theoretical correction factor on \Dnu affects the seismic radii in the luminous giant regime.

\section{Modelling \Fdnu and \Dnu}\label{sec:model methods}

From the cluster and the high-$\alpha$ sequence analysis, there is clear evidence that the breakdown of the scaling relations in the luminous giant regime could arise from issues in both \numax and \Dnu. Based on our results, \Dnu contributes more to this breakdown, but the robustness of this conclusion is limited by small sample size and large uncertainties. The source of the breakdown in the luminous giants may originate from observation, theory, or some combination of both. We focus here on theoretical modelling. 

\Fdnu is defined as the scale factor between the mean stellar density and the theoretical frequency spacings. There are two broad families of solutions to the luminous giant problem.  One is that the theoretically predicted frequency spacings are in error due to defects in our assumptions about the input physics. Another is that we are not correctly mapping the theoretical spectra onto the observed one. We refer to the former issue as model physics, and the latter issue model treatment. We choose to focus on model treatment instead of model input physics. \citet{2023MNRAS.525.5540Z} points to atmospheric and convective assumptions as a potential source of issues in the luminous giant regime. High luminosity stars have tenuous atmospheres, so therefore adopting a gray model atmosphere may not be appropriate assumption in this domain. However, we lack a compelling quantitative alternative to these assumptions. It is therefore reasonable to seek other explanations within the current physical framework to address the breakdown of the scaling relations. 

In this work, we test two different ways models can be applied to observations. The first avenue we explore addresses how calibrating the models to observed properties affects the \Fdnu correction term. In general, models of evolved stars are not calibrated to match the observed proprieties of the star. By using the mixing length to calibrate these models in radius-\Teff space, we hope to determine whether calibration affects the \Fdnu term. Second, theory predicts a wide range of oscillation modes, but not all of these modes are visible in the observed frequency spectrum. We therefore select which modes are used to compute a synthetic \Dnu term according to the number of modes we expect to observe in a star. 

\subsection{Base MESA Grid Parameters}
The synthetic frequency spectrum are computed with the GYRE stellar oscillation code using MESA stellar evolution models as inputs \citep{2011ApJS..192....3P, 2013ApJS..208....4P, 2015ApJS..220...15P, 2018ApJS..234...34P, 2019ApJS..243...10P, 2022arXiv220803651J, 2013MNRAS.435.3406T}. We use MESA version 15140. The free parameters of the stellar evolution grid include mass, metallicity, and mixing length. The quantities used for these parameters are given in Table \ref{tab: MESA parameters}.

The MESA models use a grey Eddingtion $T-\tau$ relationship with a variable opacity throughout the atmosphere. We adopt a solar mixture from \citet{1998SSRv...85..161G} (hereafter GS98) and use atomic opacity tables from GS98. The molecular opacities are given by \citet{2005ApJ...623..585F} (FA05). We use the mixing length prescription of convection from \citet{1968pss..book.....C}. The models do not consider rotation or mass loss. We use the default equations of state and nuclear reaction rates inherent to MESA. Convective overshoot is considered via the exponential formalism given in \citet{2000A&A...360..952H} where the diffusion coefficient in the overshoot region is given by:

\begin{equation}
    D_{ov} = D_0 exp(\frac{-2 z}{f H_p}),
\end{equation}

where $D_0$ is the convective coefficient near the base of the convective zone, $z$ is the distance from the edge of the convective zone, $H_p$ is the pressure scale height, and $f$ is some fraction of the pressure scale height. A parameter $f_0$ constrains $D_0$. We adopt values of $f = 0.0014$ and $f_0 = 0.004$.

The MESA stellar evolution models are computed with twice the default spatial and temporal resolution. The inlist for the base model grid is provided in Appendix I.

\begin{table}
\footnotesize
    \centering
    \begin{tabular}{c c}
         Parameter & Values  \\
         \hline
         \aml & $\{1.4 - 2.3: 0.3\}$ \\ 
         $[Fe/H]$ & $\{-0.5 - 0.5 : 0.1\}$ \\ 
          $M_*$ & $\{0.8 - 1.5: 0.1\}$, $1.8, \ 2.0$ \\ 
        
        \hline
    
    \end{tabular}
    
    \caption{MESA free parameters}
    \label{tab: MESA parameters}
\end{table}

Our model grid does not consider alpha-element enhancement. Therefore, to include the effect of alpha-abundances on the position of a star in the HR-diagram, we correct the metallicity look-up for alpha-enhancement using the Salaris approximation. The Salaris approximation adjusts the metallicity used to look-up a model in a stellar evolution grid by the following equation: 

\begin{equation}
    [Fe/H] = [Fe/H]_{obs}
 + \log{(0.638 * 10^{[\alpha/Fe]_{obs}} + 0.382)}
 \end{equation}
 \citep{1993ApJ...414..580S}.

\subsection{GYRE Calculation}
The MESA stellar structures are used as inputs into the stellar oscillation code, GYRE - which computes synthetic frequency spectra \citep{2013MNRAS.435.3406T}. We only compute radial $\ell = 0$ modes in the synthetic frequency spectra to reduce computational expense. 

We compute all of the GYRE synthetic spectra assuming the non-adiabatic assumption. It is common for studies to take the adiabatic assumption when computing the synthetic frequency spectrum. However, \citet{2023MNRAS.525.5540Z} demonstrates that stars in the luminous giant regime experience substantial departures from adiabaticity and become super-adiabatic. This effect is especially pronounced in the upper regions of the stellar atmosphere - which are a substantial component of the acoustic cavity that GYRE considers to compute the synthetic frequency spectrum. Therefore non-adiabatic effects are likely to play a significant role on the upper giant branch. \citet{2023MNRAS.525.5540Z} shows improved agreement between seismic and parallactic radii in the luminous giant regime by assuming non-adiabaticity. From these synthetic spectra we are able to measure \Dnu via multiple techniques discussed in Section \ref{sec: methods, freq spec measurement}.

\subsection{Model Interpretation}

In this section, we detail how we compute the theoretical \Dnu and \Fdnu correction from the models and how we calibrate our models to the observational data using the mixing length parameter. 

\subsubsection{Mixing Length Calibration}\label{sec: mixing length calbration}


The mixing length parameter, \aml, describes the efficiency of convection, which affects the radii of stellar models with deep surface convection zones \citep{1968pss..book.....C}. As this is not know from first principles, the Sun is commonly used as a calibration point, with the mixing length set such that solar models have the correct radius and luminosity at the solar age. However, there is no reason why this should be a universal constant. Theorists compare predicted and actual frequency spectra at fixed mass and \numax, or fixed mass and radius - which can be adjusted by adopting a different \aml. Thereby, using a fixed mixing length may neglect an important piece of physics in our models and inadequately map between observed frequency spacings and mean density. 

\citet{2017ApJ...840...17T} found a metallicity-dependent mixing length relationship for stars on the RGB. Other work including \citet{2018ApJ...856...10J} and \citet{2023Galax..11...75J} have used \aml to calibrate stellar models to observed properties. The latter work includes a discussion of potential relationships between fundamental stellar properties such as age, metallicity, and age to \aml. Here, we adopt the \citet{2017ApJ...840...17T} approach to ensure that our models align with the observed stellar \Teff and radii. We use APOGEE spectroscopic abundances and corrected seismic masses from APOKASC3 to determine the expected \Teff of the star as a function of radius. We only use stars with radii $\leq 20 R_{\odot}$ to calibrate the mixing length - metallicity relationship. For each star, we solve for the mixing length that yields the best agreement between the predicted and actual \Teff by searching in a grid of models with \aml between 1.4 to 2.3. For mixing lengths between 1.4 to 2.3 we compute the reduced $\chi ^2$ with the observed and expected temperature and find the global minimum. We then fit a linear function to the calibrated mixing lengths of the full sample as a function of the Salaris corrected stellar metallicities. This relation is used to determine a star's calibrated mixing length rather than individual calibrations.

\begin{figure*}
    \centering
    \includegraphics[width=\textwidth]{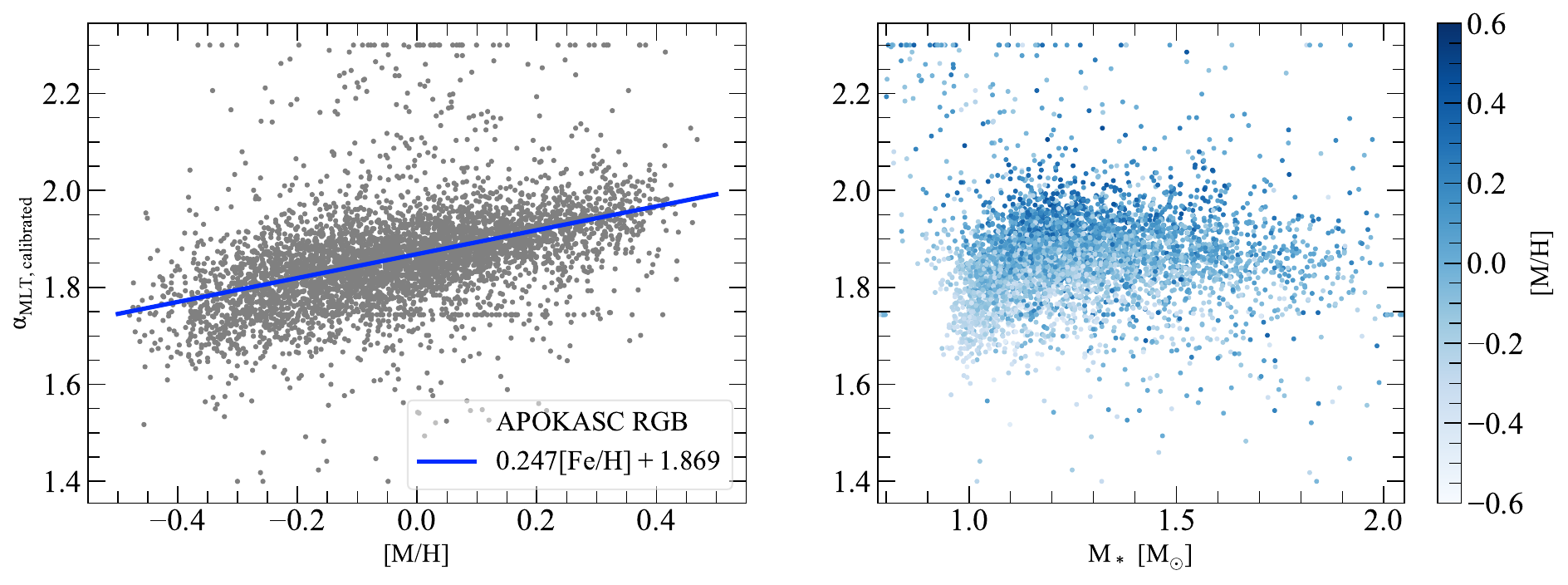}
    \caption{Calibrated mixing length relationships. Left Panel: Calibrated mixing lengths as a function of [Fe/H]. The points in grey are the individual calibrations for RGB stars in APOKASC3. The plotted [Fe/H] values are the Salaris corrected values. The blue line is the final mixing length - metallicity relationship used for determining the seismic properties from the model grid. Right Panel: Calibrated mixing length as a function of stellar mass in solar masses. The color coding indicates the Salaris-corrected metallicity for RGB stars in APOKASC3. See section \ref{sec: mixing length calbration} for discussion of this figure.}
    \label{fig: calibration relationshops}
\end{figure*}

 Similarly to \citet{2017ApJ...840...17T}, we find no correlation between temperature offsets or calibrated mixing lengths with mass, as shown in the rightmost panel of Fig. \ref{fig: calibration relationshops}. In this figure, there seems to be some correlation between the calibrated mixing length and mass for low mass stars then a plateau in higher mass ones. This however is a bias where low mass stars tend to be preferentially metal-poor.

\begin{figure}
    \centering
    \includegraphics[width=0.5\textwidth]{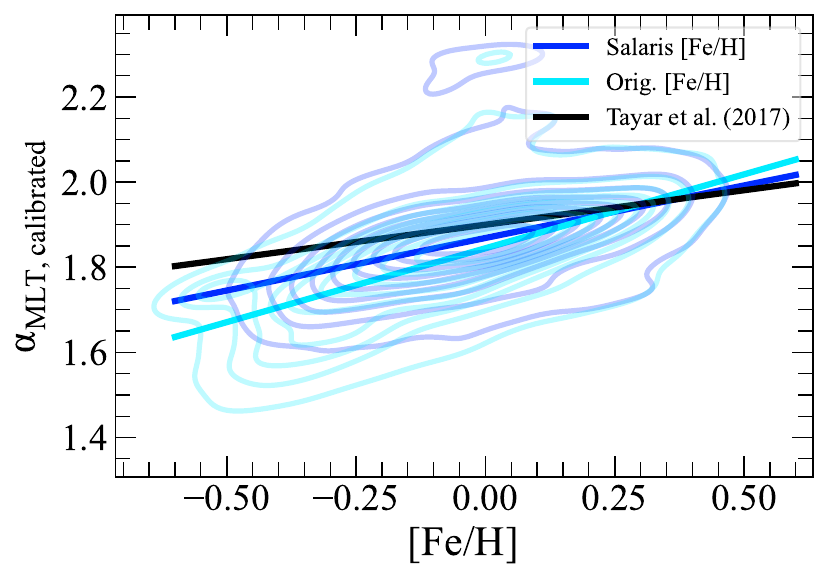}
    \caption{Calibrated mixing length as a function of [Fe/H] for different methods. The different color lines show the Salaris-corrected [Fe/H] - calibrated \aml relationship (dark blue), the original [Fe/H] - calibrated \aml relationship (light blue), and the calibrated \aml relationship determined by \citet{2017ApJ...840...17T} using YREC stellar evolution models (black). The contour lines for the Salaris-corrected and original [Fe/H] relationships indicate the density of APOKASC3 points.}
    \label{fig: calibration comparisons}
\end{figure}

The final result of this calibration is shown in Fig. \ref{fig: calibration relationshops}. We find that the calibrated relationship we get is consistent with the relationship found by \citet{2017ApJ...840...17T} shown in Fig. \ref{fig: calibration comparisons} along with a calibration relationship using the $[\alpha/Fe]$ uncorrected metallicity.

\subsubsection{Large Frequency Separation Measurement}\label{sec: methods, freq spec measurement}
For red giants, the GYRE calculation typically provides frequency measurements for $\approx \ 25-30$ radial modes. Observers collapse the frequency spectrum into a single figure of merit, \Dnu. There are several methods used to measure \Dnu in the observed frequency spectrum, each of which may yield slightly different values. We therefore mirror this exercise and apply different techniques to measure \Dnu in the theoretical frequency spectrum. The choice of which theoretical frequencies to use, and their weights will affect the mapping from theory to data. We test two methods of measuring \Dnu, or mode selection, in the theoretical frequency spectrum: (i) using all available modes computed by GYRE, and (ii) using only a subset of modes from the synthetic frequency spectrum that is similar to the number of modes that would be observationally accessible. Most current techniques use this subset or a similar subset of modes to compute \Dnu (e.g. \citet{2009CoAst.160...74H, 2016ApJ...833L..13G, 2009A&A...508..877M}) While there are many ways that \Dnu is measured observationaly, we elect to test these two methods as we need a system that is robust to the choice of observational method. Further, \citet{2018ApJS..239...32P} and \citet{2024arXiv241000102P} include discussions of the effect of frequency measurement and mode averaging techniques on systematic uncertainties in the \Dnu measurement. Taking the uncertainty measure as scatter between the different pipeline measurements, \citet{2024arXiv241000102P} shows that for the lower RGB gold sample \numax and \Dnu are measured to within $1\%$ across the different techniques. In the luminous giant regime, there is greater deviations between the techniques at the $\approx 2 - 3\%$ level between the different pipelines. It is important to note that these deviations in the luminous giant regime contribute to greater uncertainty in the seismic radii of the luminous giants and therefore greater uncertainty in our model-conclusions.

The first method we apply uses a weighted average applied to all of the modes. We apply a Gaussian envelope centered on \numax with a width defined in \citet{2012A&A...537A..30M} as:

\begin{equation}
    \delta \nu_{env} = \alpha \nu_{max}^\beta .
\end{equation}

Where $\alpha = 0.66$ and $\beta = 0.88$. The exact choices for the values defining the Gaussian envelope will vary between studies, but this technique is common in studies using an auto-correlation function to measure \Dnu \citep{2009CoAst.160...74H, 2016ApJ...833L..13G, 2009A&A...508..877M}. We then multiply the value  of the Gaussian envelope at the midpoint between two adjacent radial modes to the frequency separation between those modes. The summation of the Gaussian weighted frequency separations across all of the radial modes measured by GYRE is then the measured theoretical \Dnu value. We also add an unweighted average of all of the mode separations to this analysis as a comparison point. 

\begin{figure}
    \centering
    \includegraphics[width=0.5\textwidth]{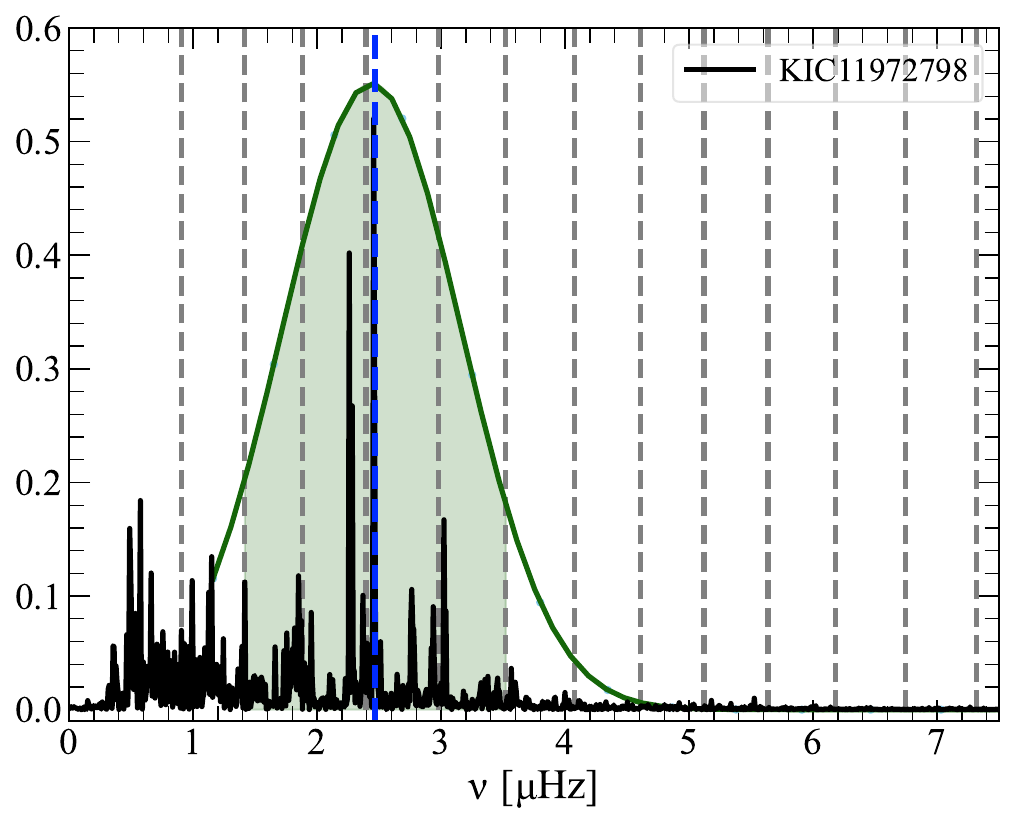}
    \caption{Example frequency spectrum measurement. In black, we have plotted the frequency spectrum for \textit{KIC11972798}. This star has $M_* = 0.9314 \pm 0.1115 M_{\odot}$ and $R_* = 36.8295 \pm 0.0535 R_{\odot}$. The dotted grey lines are the $l = 0$ mode measurements computed from GYRE. The dotted blue line is $\nu_{max} = 2.4582 \ \mu Hz$ for this star. The green Gaussian is the Gaussian weights applied to the modes computed by GYRE. The green shaded region indicates where the modes are truncated to compute \Dnu with a subset of modes. See sec. \ref{sec: methods, freq spec measurement} for details about how we measure \Dnu from the theoretical frequency spectra.}
    \label{fig: Dnu measurement}
\end{figure}

The second method we use to measure \Dnu uses only a subset of frequency measurements given by GYRE. In the luminous giant regime, we typically only observe around $4-6$ modes \citep{2013A&A...559A.137M}. By only using a subset of modes computed by GYRE, we mimic observations of luminous giant stars. The number of modes we use is given by the relation: 

\begin{equation}\label{eq:5}
    n_{env} = \frac{\delta \nu_{env}}{\Delta \nu}
\end{equation}

from \citet{2012A&A...537A..30M}. We apply the same Gaussian envelope as in the weighted average technique, but we truncate the number of modes used to compute the Gaussian weighted frequency separation according to the number of modes we expect to observe. This corresponds to two times $n_{env}$. Figure \ref{fig: Dnu measurement} illustrates the two methods of measuring \Dnu. We expect some differences between these two methods as the frequency separations differ from mode to mode. By including all of the modes, we include frequency separations that are far from the \numax value and therefore may be closer to the asymptotic limit. 

\subsection{Modelling \Fdnu Results} \label{sec: results models}

\begin{figure*}
    \centering
    \includegraphics[width=\textwidth]{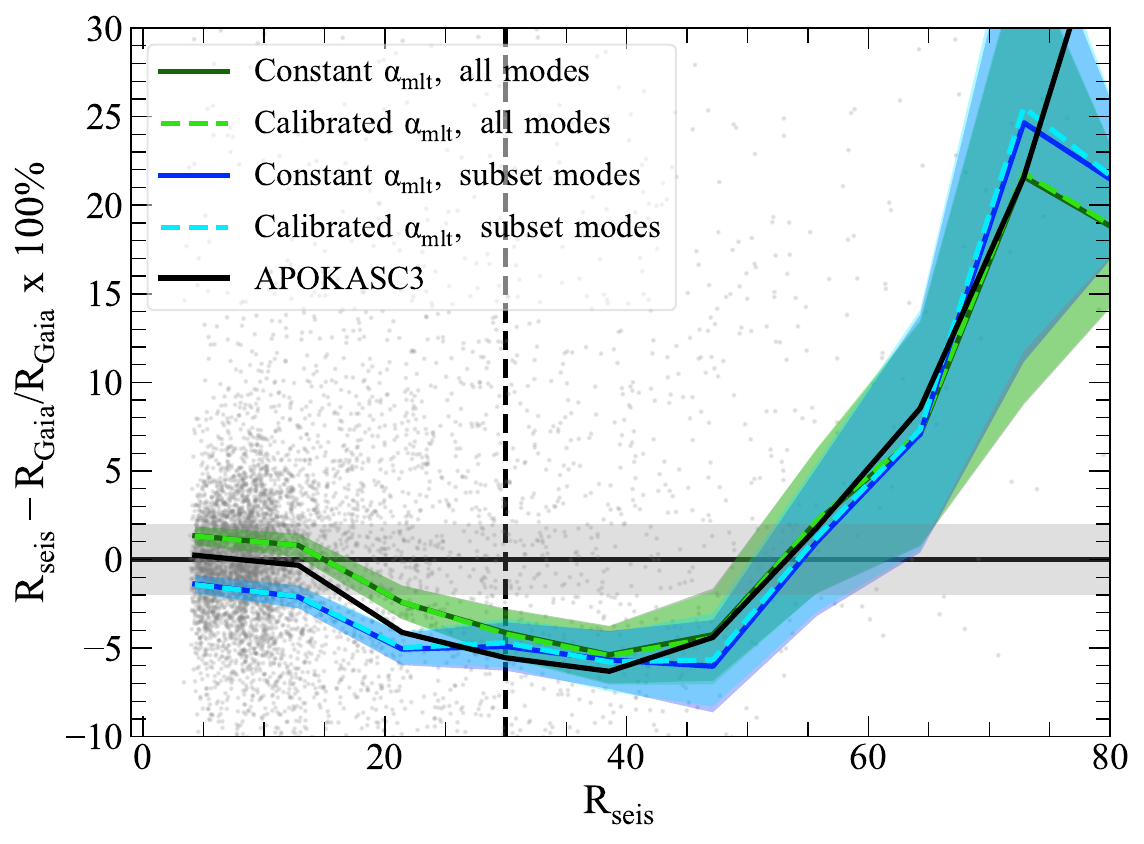}
    \caption{Fractional difference between $R_{seis}$ and $R_{gaia}$ v. $R_{seis}$ for different models. The solid colored line uses \Fdnu correction factors for uncalibrated models and the dashed lines use the \aml - $[Fe/H]$ calibration. Green lines use all of the available modes computed by GYRE to measure \Dnu in the theoretical frequency spectrum. Blue lines use a subset of modes to compute the theoretical \Dnu. See sections \ref{sec: mixing length calbration} and \ref{sec: methods, freq spec measurement} for details on the mixing length calibration and frequency spectrum measurement. The solid black line is the mass given in the APOKASC3 catalog. The black vertical dashed line and the solid black horizontal dashed lines are meant to guide the eye. The vertical black dashed line shows where we begin to consider stars luminous giants and the solid black horizontal line is where there is zero offset between the seismic radius and the parallactic radius. The grey shaded region is $\pm 2\%$ differences between seismic and parallactic radius. Refer to sec. \ref{sec: results models} for discussion of this plot.}
    \label{fig: model interp results}
\end{figure*}

In this section, we summarize how the agreement between seismic radius and parallactic radius is affected by mixing length calibration and \Dnu measurement technique. Figure \ref{fig: model interp results} summarizes our findings for both approaches. 

\begin{figure}
    \centering
    \includegraphics[width=0.5\textwidth]{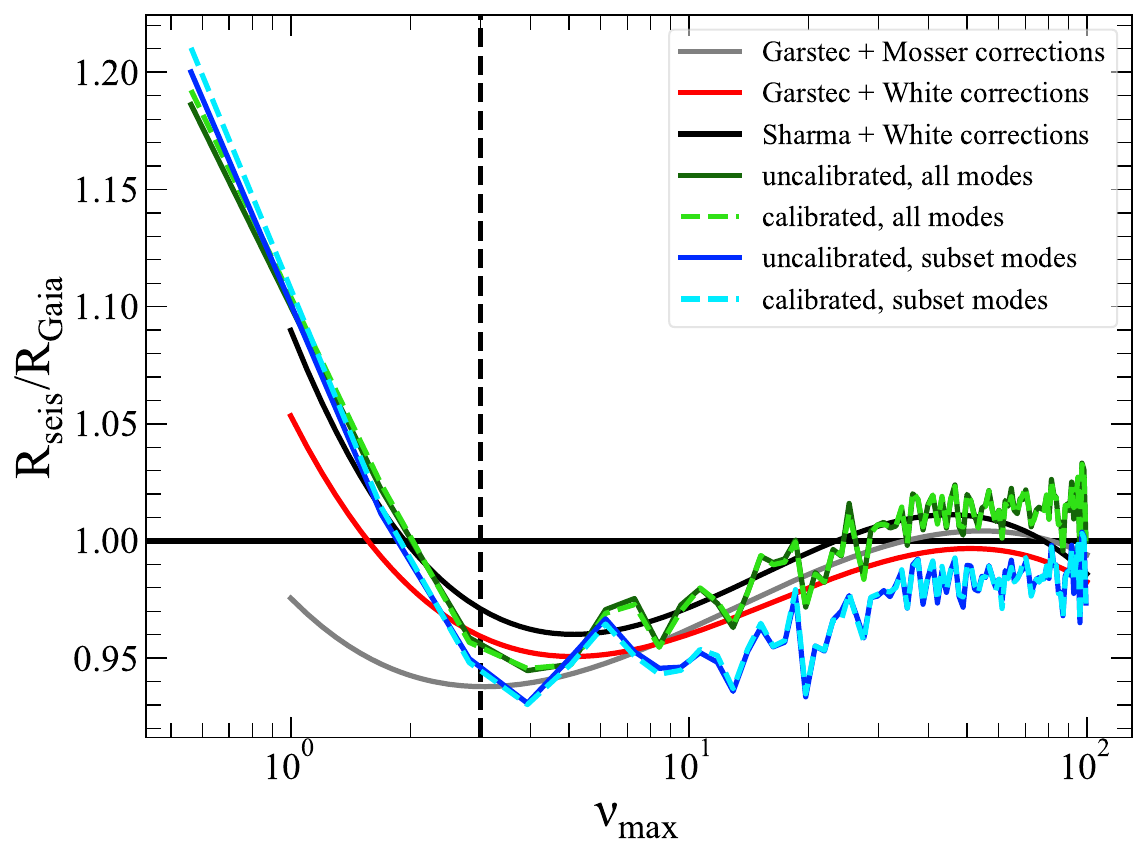}
    \caption{Comparison between calibration scales used in this paper and the APOKASC3 catalog. The axes are the seismic radius computed with different correction terms over the \textit{Gaia} radius versus \numax. Results from this study are given by the dark blue, light blue, dark green, and light green lines. They are compared to calibration polynomials available in the APOKASC3 catalog given by the red, grey, and black lines. }
    \label{fig: seismic calibrations}
\end{figure}

\begin{figure*}
    \centering
    \includegraphics[width=1.0\textwidth]{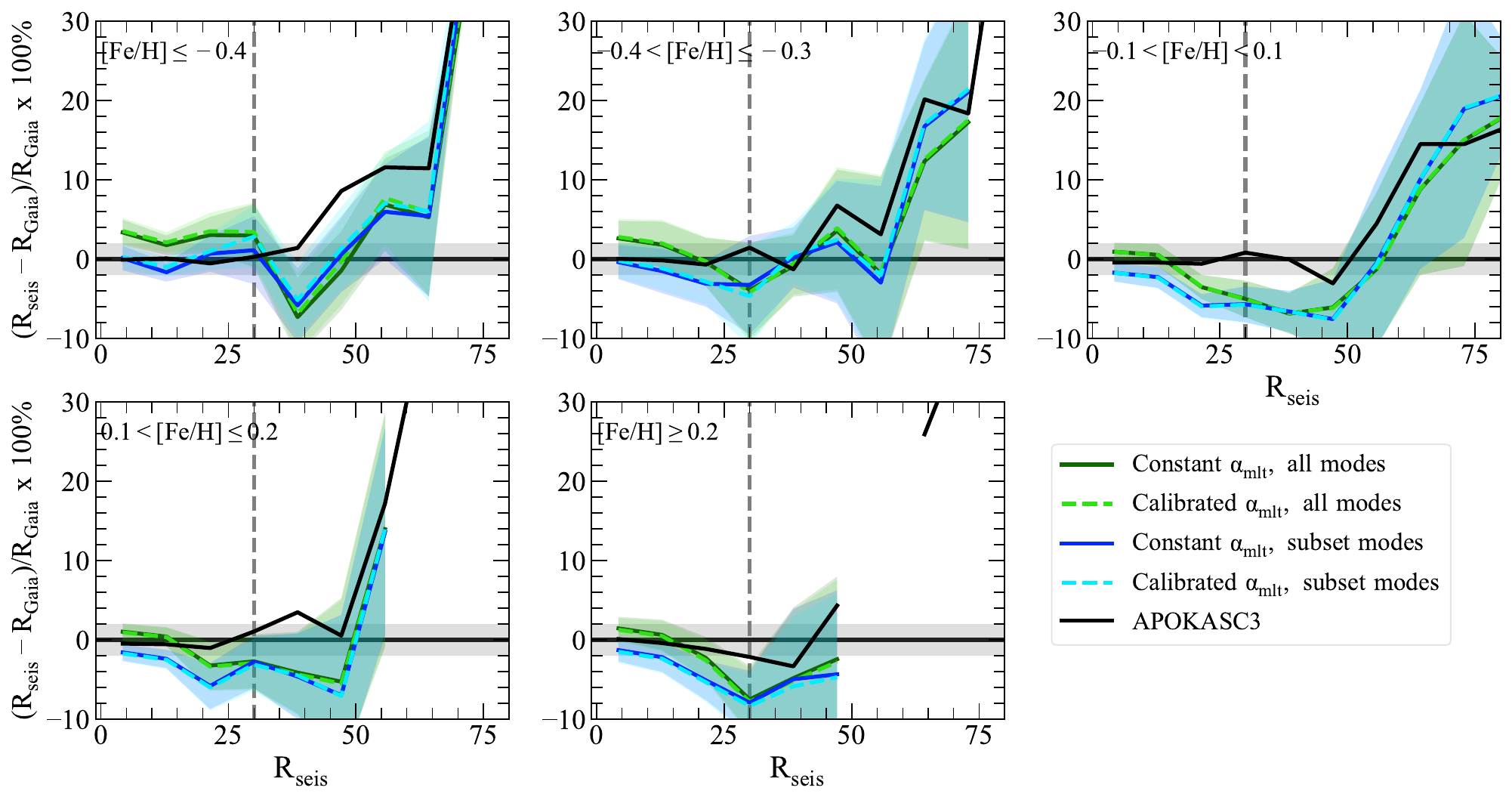}
    \caption{Fractional difference between $R_{seis}$ and $R_{gaia}$ v. $R_{seis}$ for different models. The plot follows the same conventions as \ref{fig: model interp results}. However, this plot has been broken into sub-panels for different metallicity bins. }
    \label{fig: model interp results metallicity}
\end{figure*}

\subsubsection{Mixing Length Calibration} \label{sec: mixing length results}

In general, the predicted locus of stellar models need not agree with the observed one.  In this case, if we compare models with data at a fixed \numax, or at a fixed radius, we would get different results.  Fortunately, calibrating model tracks in the HR-diagram using a metallicity dependent mixing length relation has little impact on the agreement between parallactic radii and seismic radii in the APOKASC3 sample. As shown in Fig. \ref{fig: model interp results} and Fig. \ref{fig: seismic calibrations}, the differences between mixing length calibrated and uncalibrated seismic radii are very small. Averaged across the RGB, the difference between the calibrated and uncalibrated radii is $0.1 \%$. In the luminous giant regime the mean difference between the calibrated and uncalibrated cases is $0.2\%$ and $0.3\%$ for radii measured using all of the modes and a subset of modes, respectively. Fig. \ref{fig: model interp results} and Fig. \ref{fig: model interp results metallicity} shows the binned fractional difference between the seismic radii and the parallactic radii as a function of seismic radius. The dotted and dashed lines are the uncalibrated and calibrated models respectively. We see in the full sample that the uncalibrated and calibrated models agree to large seismic radius. 

Figure \ref{fig: seismic calibrations} compares the seismic radii computed in this study over the \textit{Gaia} radii to calibration systems included in the APOKASC3 catalog \citep{2024arXiv241000102P}. One can use the seismic radius over the parallactic radius to define a correction term, \Fnmax, as  a function of \numax to correct seismic radius onto the \textit{Gaia} radius scale. In this figure, we can see that the topology of our calibration system tracks with the topology of the APOKASC3 systems as a function of \numax. Differences between our calibration system and the other systems in APOKASC3 can be traced to different physical assumptions and model choice. Specifically, different stellar evolution models will yield slightly different answers and we adopt the non-adiabatic assumption which is not a common choice among models. An advantage of our system is that it extends into the regime of $\nu_{max} \leq 1$, whereas the other systems do not. 

The full APOKASC3 sample is mostly solar-like metallicity stars. Since the uncalibrated mixing length is set to the solar-calibrated mixing length, the difference between calibrated and uncalibrated mixing lengths in the full sample is expected to be small. Hence, we also look at the agreement between parallactic and seismic radii at different metallicities in Fig. \ref{fig: model interp results metallicity}. Despite the metallicity-dependence for the calibrated mixing length, we still see similar differences between the calibrated and uncalibrated models as in the full sample. 

\begin{figure}
    \centering
    \includegraphics[width=0.48\textwidth]{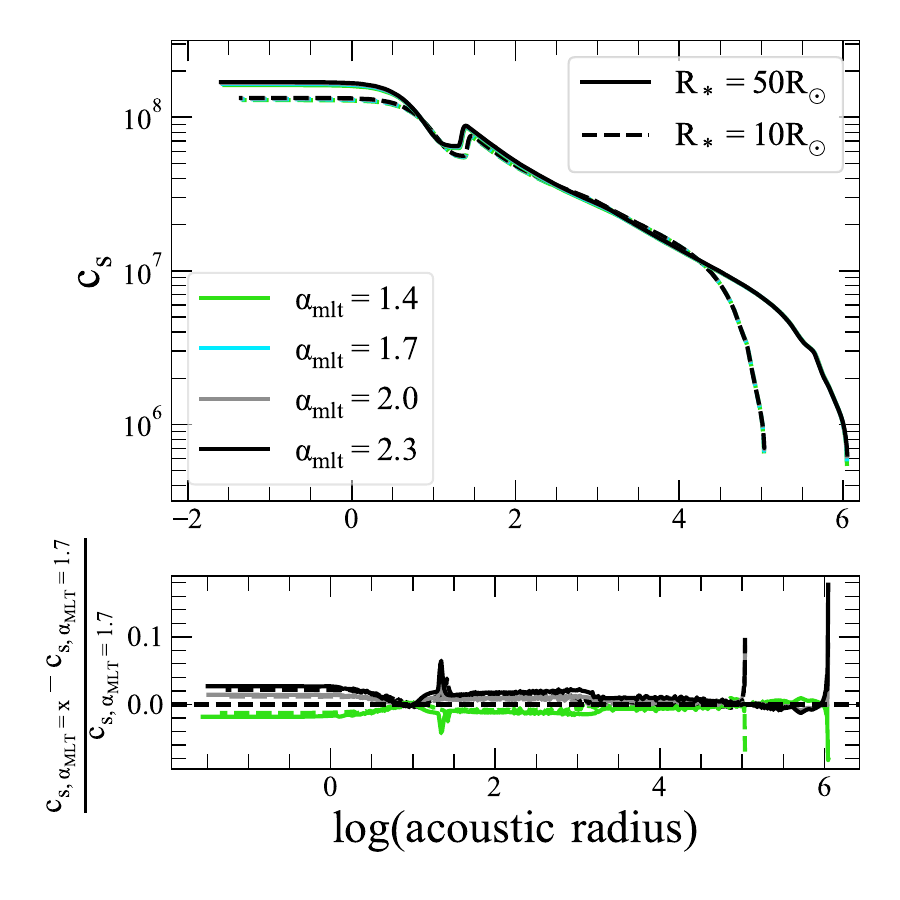}
    \caption{Sound speed vs. acoustic radius for models with different radii and mixing lengths. The solid and dashed lines refer to stars with $50 R_{\odot}$ and $10 R_{\odot}$, respectively. The different color lines indicate different mixing lengths in the model. The bottom panel shows the fractional difference in the sound speed profiles compared to the sound speed profile for a mixing length of $1.7$ as a function of the acoustic radius for the different radii stars. Refer to sec. \ref{sec: mixing length results} for plot interpretation.}
    \label{fig: sound speed profiles}
\end{figure}

To illustrate why mixing length has little impact on the inferred seismic radius, we plot the sound speed profiles for different mixing lengths and stellar radii as a function of the acoustic radius in Fig. \ref{fig: sound speed profiles}. We include two stellar radii in this figure: $10 R_{\odot}$ as a lower RGB star and $50 R_{\odot}$ as a luminous giant star. All of the models in this figure are solar metallicity. We quantify the difference between sound speed profiles in the lower panel of Fig. \ref{fig: sound speed profiles}, which plots the fractional difference between a given mixing length's sound speed profile compared to the $\alpha_{MLT} = 1.7$ sound speed profiles. We see that the different mixing lengths produce sound speed profiles that are similar to one another at the $2 - 6 \%$ level in the interior of the star. In general, we see the greatest differences between the sound speed profiles deep in the star at low acoustic radius. As we move towards the outer regions of the star, the differences between the sound speed profiles shrink to the sub percent level. 

Despite the difference between the sound speed profiles deep in the star, we do not see an impact on the \Fdnu correction factor, and therefore the seismic radii due to mixing length calibration. This is because the deepest regions of the star have little impact on the frequency spacings. The outer envelope of the star - where the differences between the sound speed profiles are sub-percent, has the largest impact on the frequency spectrum. If the mixing length affected the sound speed profile in outer regions of the star, then we may expect to see some differences between the mixing length calibrated and uncalibrated seismic radii. 

\begin{figure}
    \centering
    \includegraphics[width=0.48\textwidth]{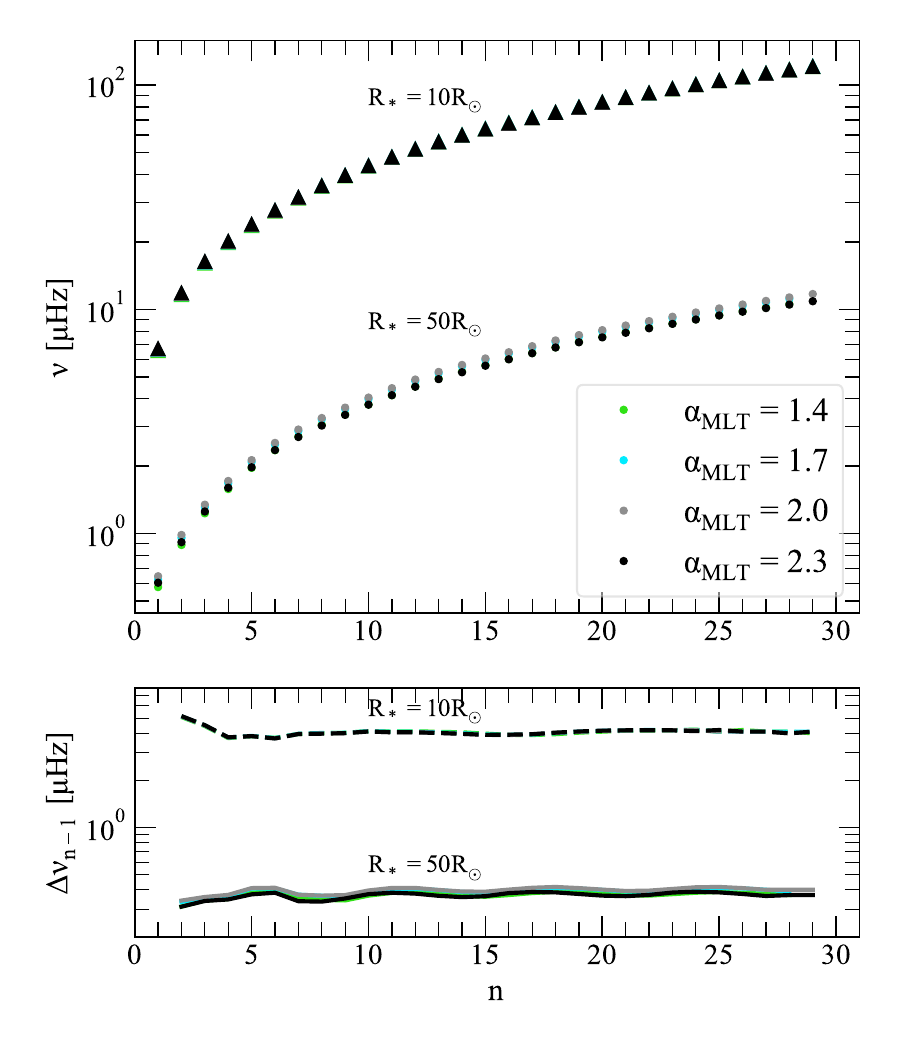}
    \caption{Frequency spectra for models of different stellar radius and mixing length. Top panel: The y-axis of this figure is the frequency of a given mode on the x-axis. The triangle points are measurements for a model with a radius of $10 R_{\odot}$. The circle points have a radius of $50 R_{\odot}$. The point colors indicate the mixing length of the model. Bottom panel: Frequency separations between adjacent modes plotted as a function of mode. The dotted lines are models with $R_* = 10 {R_{\odot}}$ and the solid lines are $R_* = 50 R_{\odot}$. The colors of the lines are different mixing lengths and have the same colors as the legend in the top panel. Section \ref{sec: mixing length results} discusses the results of this figure.}
    \label{fig: freq spec mixing length}
\end{figure}

The similarities in sound speed profiles indicate that the frequency spacings are insensitive to the mixing length parameter. Figure \ref{fig: freq spec mixing length} shows the calculated frequencies and the frequency separations for several mixing lengths as a function of mode. We see in this figure that there are some small changes in the absolute frequencies, but there is little change in the frequency spacings as the mixing length changes. There are slightly larger differences between the frequency spacings for the $50 R_{\odot}$ star . This explains why we see some deviation between mixing length calibrated and uncalibrated models at the largest radii in Fig. \ref{fig: model interp results}.

\subsubsection{Mode Selection}\label{sec:Freq measurement results}

\begin{figure}
    \centering
    \includegraphics[width=0.5\textwidth]{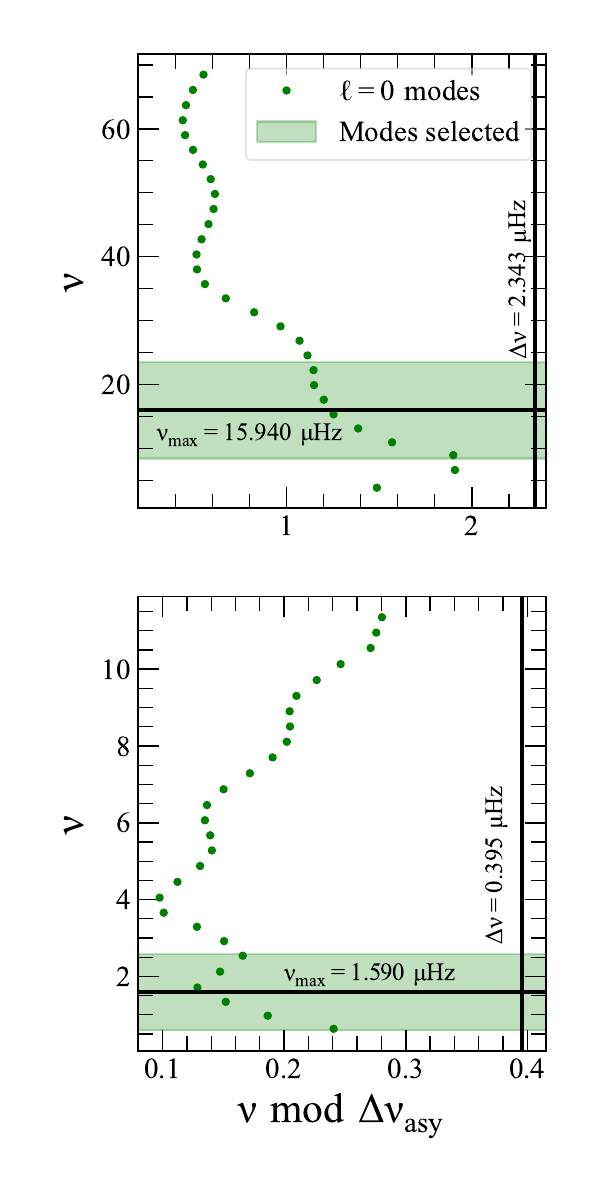}
    \caption{Echelle diagram for synthetic frequency spectra. Top panel: Synthetic frequency spectrum for a $1 M_{\odot}$, $15 R_{\odot}$ star. Bottom panel: Synthetic frequency spectrum for a $1 M_{\odot}$, $50 R_{\odot}$ star. The green shaded region is the width of the Gaussian envelope used to select modes to compute \Dnu. }
    \label{fig: echelle fig}
\end{figure}

In addition to testing the effect of mixing length calibration on the seismic radii, we investigate how measuring \Dnu in the synthetic frequency spectrum affects the \Fdnu correction factor and in turn the seismic radius. In Fig. \ref{fig: model interp results}, we show how different \Dnu measurement techniques affect the fractional difference between the parallactic radius and the seismic radius as a function of seismic radius. In Fig. \ref{fig: model interp results}, we can isolate the different measurement techniques by focusing on the mixing length uncalibrated models. On the lower RGB, the weighted average technique that uses all of the modes produces radii that are over-inflated relative to the \textit{Gaia} radii and the technique that uses a subset of modes in the weighted average produces stars that are too small relative the \textit{Gaia} radii. In the luminous giant regime, the two techniques track one another up to a radius of $\approx 70 R_{\odot}$ where the weighted average of all of the modes produces slightly more consistent results. Finally, Fig. \ref{fig: echelle fig} demonstrates why using a subset of modes versus all of the modes to compute \Dnu produces different results. In this figure, we provide two example synthetic echelle diagrams, one for a solar mass, $15 R_{\odot}$ star and one for a $50 R_{\odot}$ star. We can see in both cases that at large frequencies, the separations drift from the asymptotic \Dnu value. When we use all of the modes to compute \Dnu in the synthetic spectra, we capture this behavior, whereas when we use an observationally motivated subset of modes, we do not. 

From this study alone, we don’t see a statistically significant difference between the seismic radii predicted from the different \Dnu measurement techniques. While the method that uses all of the modes produces radii that are closer to the \textit{Gaia} radii in the luminous giant regime, we cannot conclude with a high degree of certainty that this technique is better. The similarities between the results from the two techniques indicates that this particular avenue cannot solve the luminous giant problem. We therefore still recommend the synthetic \Dnu value be measured with a subset of modes as this technique is observationally motivated. Further, as shown in \ref {fig: echelle fig}, this technique excludes modes at high frequencies that show larger deviations from the asymptotic frequency spacings.

\section{Summary \& Future Work}\label{sec: conclusions}

In this paper, we investigate the efficacy of the asteroseismic scaling relations in the luminous giant regime, where there is an observed breakdown. The frequency of maximum power \numax is related to the surface gravity, while the frequency spacing \Dnu is related to the mean density. In principle, the deviations that we see could reflect issues with either or both of these terms. To test which of the asteroseismic parameters, \numax or \Dnu may be causing the breakdown, we use open clusters and the high-$\alpha$ sequence to probe the single-parameter asteroseismic scaling relations. We use these stars as they have known masses and radii to compare against. From this analysis, we see that both the single scaling relations are in good agreement with seismic-independent parameters on the lower giant branch. However, we see some statistically insignificant departures in the luminous giant regime for the \Dnu relation. The reliability of the \numax scaling relation is important, and implies that a solar scaled \numax may be a reliable gauge of surface gravity at the tip of the giant branch. 

For \Dnu, unlike \numax, we have a well-posed theoretical framework for addressing possible issues in the scaling relations. We therefore test how changing the theoretical \Fdnu correction factor affects the agreement between seismic radii and parallactic radii. We test how different model interpretations, namely choice in the mixing length parameter and how \Dnu is measured in the synthetic frequency spectrum affects the \Fdnu factor. We find that calibrating the models with a metallicity-dependent mixing length has little affect on the inferred seismic radius. The \Dnu measurement technique does have a small effect on the agreement between seismic radii and parallactic radii. Our main conclusions are summarized below: 

\begin{itemize}

\item Throughout the literature, the asteroseismic scaling relations are shown to be robust on the lower RGB. Our results fortify this conclusion. We demonstrate that seismic properties are in good agreement with seismic-independent ones in the full APOKASC3 sample, the open cluster sample, and in the high-$\alpha$ sequence. Further, with zero-point corrections, our results in the full APOKASC3 sample are in good agreement on the lower RGB regardless of the method used to interpret the theoretical frequency spectrum. 

\item Previous work has demonstrated a breakdown in the joint scaling relations in the luminous giant regime. Typically, this has been ascribed to an ill-defined \Dnu measurement for these stars. It is easier to measure a power excess, i.e. \numax in the luminous RGB domain than it is to measure frequency separations due to a small number of observable modes. Using the Milky Way's high-$\alpha$ sequence, we demonstrate that the \numax-only relationship is a reliable figure of merit excluding the most evolved RGB stars. \Dnu however appears to produce more discrepant results and mirrors trends in the joint scaling relation. The robustness of our conclusion about the \Dnu-only scaling relation is limited by its statistical significance. We find that both \numax and \Dnu are likely to play some role in the breakdown of the full asteroseismic scaling relation in the luminous RGB regime to different extents. This warrants caution for future studies that use single-parameter scaling relations in extending into the luminous giant domain. 

\item We demonstrate that the method used to interpret the theoretical frequency spectra and how the theoretical frequency spectra are compared to observations has an effect on the inferred seismic properties. Though, the differences between the techniques are small. Measuring \numax and \Dnu from theoretical frequency spectra and matching these spectra to observations is subjective with many studies adopting different methods. Further work should be done to determine the optimal treatment of theoretical frequency spectra as it would be advantageous for a uniform approach to be adopted in the field. 

\item Additional work is needed to determine the physical source of the breakdown in the astreroseismic scaling relationships in the luminous RGB regime. There are several assumptions in stellar evolution theory that are not reflective of the conditions in luminous RGB stars. For example, convection in luminous RGB stars may deviate from the assumptions posed by mixing-length theory as their convective envelopes are extended and turbulent. This approach would likely require computationally expensive 3-dimensional hydrodynamic simulations of convection. A more immediately tractable avenue for exploring the breakdown in the luminous RGB would be investigating the effect of atmospheric assumptions on the inferred seismic properties. Luminous RGB stars may not be well represented by the solar-like atmospheres typically assumed in stellar evolution models. An investigation of this kind would involve invoking different equations of state for the stellar atmosphere, solving for radiative transfer, and propagating the new atmosphere through the stellar evolution models. This avenue is particularly enticing as the stellar atmosphere of luminous giants has a substantial impact on the frequency spectrum. The bulk of the acoustic cavity is in the stellar atmosphere of these stars, so one may expect changes in the atmospheric assumptions will affect the frequency spectra. 

\end{itemize}

Our results have implications for the implementation of asteroseismology in future work. Most pressingly, we demonstrate the need for further investigation of the single-parameter scaling relations as neither of the single-scaling relations produces consistent, reliable results on the upper giant branch. This impacts studies that plan to use a single asteroseismic measurement to determine radius or mass such as the \citet{2022AJ....164..135H} sample where only \numax is measured. In these cases, the single-parameter scaling relations will need to be carefully calibrated before being applied to large population studies. For example, studies that use asteroseismic parameters to train neural networks to determine stellar properties from spectra should take caution when training their models on luminous giant stars. While the single-parameter scaling relations should be treated with caution on the upper RGB, we do find the single-parameter scaling relations work well on the lower RGB. 

We also demonstrate the need for improved modelling to address issues with the theoretical \Fdnu correction factor. While none of the model interpretations we investigated solve the breakdown in the scaling relations in the luminous giant regime, more advanced physics may make headway. Specifically, advances in the modelling of convective physics and different treatments of the model atmosphere may have a substantial impact on the theoretical \Fdnu correction factors. 

\section*{Acknowledgements}

AA thanks Jennifer Johnson and the Ohio State University stars research group for feedback throughout the course of this work. AA also thanks M. Joyce for helpful guidance on the MESA stellar evolution model. 
MHP acknowledges support from NASA grant 80NSSC24K0637. M.V acknowledges support from NASA grant 80NSSC18K1582 and funding from the European Research Council (ERC) under the European Union’s Horizon 2020 research and innovation programme (Grant agreement No. 101019653).

This work has made use of data from the European Space Agency (ESA) mission
{\it Gaia} (\url{https://www.cosmos.esa.int/gaia}), processed by the {\it Gaia}
Data Processing and Analysis Consortium (DPAC,
\url{https://www.cosmos.esa.int/web/gaia/dpac/consortium}). Funding for the DPAC
has been provided by national institutions, in particular the institutions
participating in the {\it Gaia} Multilateral Agreement.

This paper includes data collected by the Kepler mission and obtained from the MAST data archive at the Space Telescope Science Institute (STScI). Funding for the Kepler mission is provided by the NASA Science Mission Directorate. STScI is operated by the Association of Universities for Research in Astronomy, Inc., under NASA contract NAS 5–26555.

Funding for the Sloan Digital Sky Survey IV has been provided by the Alfred P. Sloan Foundation, the U.S. Department of Energy Office of Science, and the Participating Institutions. 

SDSS-IV acknowledges support and resources from the Center for High Performance Computing  at the University of Utah. The SDSS website is www.sdss4.org.

SDSS-IV is managed by the 
Astrophysical Research Consortium for the Participating Institutions of the SDSS Collaboration including the Brazilian Participation Group, the Carnegie Institution for Science, Carnegie Mellon University, Center for Astrophysics | Harvard \& Smithsonian, the Chilean Participation Group, the French Participation Group, Instituto de Astrof\'isica de Canarias, The Johns Hopkins University, Kavli Institute for the Physics and Mathematics of the Universe (IPMU) / University of Tokyo, the Korean Participation Group, Lawrence Berkeley National Laboratory, Leibniz Institut f\"ur Astrophysik Potsdam (AIP),  Max-Planck-Institut f\"ur Astronomie (MPIA Heidelberg), Max-Planck-Institut f\"ur Astrophysik (MPA Garching), Max-Planck-Institut f\"ur Extraterrestrische Physik (MPE), National Astronomical Observatories of China, New Mexico State University, New York University, University of Notre Dame, Observat\'ario Nacional / MCTI, The Ohio State University, Pennsylvania State University, Shanghai Astronomical Observatory, United Kingdom Participation Group, Universidad Nacional Aut\'onoma de M\'exico, University of Arizona, University of Colorado Boulder, University of Oxford, University of Portsmouth, University of Utah, University of Virginia, University of Washington, University of Wisconsin, Vanderbilt University, and Yale University.

 This work made use of Astropy:\footnote{http://www.astropy.org} a community-developed core Python package and an ecosystem of tools and resources for astronomy \citep{astropy:2013, astropy:2018, astropy:2022}. 

We would like to acknowledge the land that The Ohio State University occupies is the ancestral and contemporary territory of the Shawnee, Potawatomi, Delaware, Miami, Peoria, Seneca, Wyandotte, Ojibwe and many other Indigenous peoples. Specifically, the university resides on land ceded in the 1795 Treaty of Greeneville and the forced removal of tribes through the Indian Removal Act of 1830. As a land grant institution, we want to honor the resiliency of these tribal nations and recognize the historical contexts that has and continues to affect the Indigenous peoples of this land.

\section*{Data Availability}
The APOKASC3 catalog and \textit{Gaia} DR3 data used for this study are publicly available. Datasets for the open clusters, high-$\alpha$ sequence are hosted on https://doi.org/10.5281/zenodo.14008022.


\bibliography{main}{}
\bibliographystyle{aasjournal}



\appendix
\section{Example MESA inlist}
\begin{verbatim}
&kap
use_Type2_opacities = .true.
Zbase = 1.7d-2
kap_file_prefix = "gs98"
kap_CO_prefix = "gs98_co"
kap_lowT_prefix = lowT_fa05_gs98      
/ ! end of kap namelist

&eos
/ ! end of eos namelist

&star_job
show_log_description_at_start = .false.
load_saved_model = .true.
saved_model_name = 'final_turnoff.mod'
save_model_when_terminate = .true.
save_model_filename = 'final_rgb.mod'
write_profile_when_terminate = .true.
filename_for_profile_when_terminate = 'final_rgbprofile.data'
history_columns_file = 'history_columns_w_freq.list'              
profile_columns_file = 'profile_columns.list'
/ ! end of star_job namelist

&controls
use_dedt_form_of_energy_eqn = .true.
use_gold_tolerances = .true.
mesh_delta_coeff = 0.5
time_delta_coeff = 0.5
max_years_for_timestep = 1d6
varcontrol_target = 1d-3
max_timestep_factor = 2d0
photosphere_r_upper_limit = 1.5d2 !Terminate at RGB tip
num_trace_history_values = 2
trace_history_value_name(1) = 'rel_E_err'
trace_history_value_name(2) = 'log_rel_run_E_err'
initial_mass = 1.0
initial_z = 1.7d-2
initial_y = 0.28      
do_conv_premix = .true.
mixing_length_alpha = 1.7
overshoot_scheme(:)    = 'exponential'
overshoot_zone_type(:) = 'any'
overshoot_zone_loc(:)  = 'any'
overshoot_bdy_loc(:)   = 'any'
overshoot_f(:)  = 0.014d0
overshoot_f0(:) = 0.004d0
overshoot_D0(:) = 0
use_Ledoux_criterion = .true.      
write_pulse_data_with_profile = .true.
pulse_data_format = 'GYRE'
format_for_FGONG_data = '(1p,5(E16.9))'
add_center_point_to_pulse_data = .false. 
add_atmosphere_to_pulse_data = .true.
atm_option = 'T_tau'
atm_T_tau_relation = 'Eddington'
atm_T_tau_opacity = 'varying'
log_directory = 'LOGS_rgb'
max_num_profile_models = 12
profile_interval = 100
photo_interval = 1000
terminal_interval = 1000
cool_wind_RGB_scheme = ''
cool_wind_AGB_scheme = ''
RGB_to_AGB_wind_switch = 1d-4
Reimers_scaling_factor = 0.7d0
Blocker_scaling_factor = 0.7d0
cool_wind_full_on_T = 1d10 !K
hot_wind_full_on_T = 1.1d10 !K
      
!GYRE_output_controls !custom controls to run GYRE and MESA simultaneously
x_integer_ctrl(1) = 125 ! output GYRE info at this step interval
x_logical_ctrl(1) = .false. ! save GYRE info whenever save profile
x_integer_ctrl(2) = 100 ! max number of modes to output per call
x_logical_ctrl(2) = .true. ! output freq_spec files
x_integer_ctrl(3) = 0 ! mode 
x_integer_ctrl(4) = 1 ! order
x_logical_ctrl(3) = .false. !.false. = nad, .true. = ad
!extra profile controls
x_ctrl(3) = 0.1
/ ! end of controls namelist

&pgstar

/ ! end of pgstar namelist

\end{verbatim}

\section{GYRE inlist}
\begin{verbatim}
&model
model_type = 'EVOL'  
file = 'profile200.data.GYRE'
file_format = 'MESA' 
/

&constants                                                                                                                       
G_GRAVITY = 6.6740800000e-08                                                                                             
R_sun = 6.958e10                                                                                               
M_sun = 1.988435e33                                                                                                      
/                                                                                                                                
&mode
l = 0                     
tag = 'radial'
/
&osc
outer_bound = 'VACUUM'                                          
nonadiabatic = .TRUE.  
adiabatic = .FALSE.
tag_list = 'radial'
alpha_thm = 1
	
/
&rot
! no rotation
/
&num
diff_scheme = 'MAGNUS_GL2' 
nad_search = 'MINMOD'
restrict_roots = .FALSE.
/

&scan
grid_type = 'LINEAR' 
freq_min_units = 'NONE'  
freq_max_units = 'NONE'
freq_min = 1.0       
freq_max = 40.0        
n_freq = 2000          
tag_list = 'radial'  
/

&scan
grid_type = 'INVERSE'   
freq_min_units= 'NONE'                                   
freq_max_units = 'NONE'
freq_min = 1.0           
freq_max = 40.0           
n_freq = 7000           
tag_list = 'non-radial' 
/
&grid
w_osc = 10                                                                    
w_exp = 2                                                              
w_ctr = 10 
/
&shoot_grid 
/

&recon_grid 
/

&nad_output
summary_file = 'FILEOUT.gyre_nad.eigval.h5'
summary_file_format = 'HDF'                             
summary_item_list = 'M_star,R_star,L_star,l,n_pg,n_g,omega,freq,E,E_norm,E_p,E_g' !
detail_template = 'FILEOUT.gyre_nad.mode-%J.h5'                      		  
detail_file_format = 'HDF'                   		    
detail_item_list = 
'M_star,R_star,L_star,m,rho,p,n,l,n_p,n_g,omega,freq,E,E_norm,W,x,V,As,U,c_1,Gamma_1,
nabla_ad,delta,xi_r,xi_h,phip,dphip_dx,delS,delL,delp,delrho,delT,dE_dx,dW_dx,T,E_p,E_g'
freq_units = 'UHZ'
/

\end{verbatim}

\end{document}